\documentclass[a4paper,11pt]{article}
\pdfoutput=1 

\usepackage{jheppub} 
\usepackage{physics}
\usepackage[T1]{fontenc} 
\usepackage{appendix}
\usepackage{amsmath, amssymb, graphicx}
\usepackage{amsfonts}
\usepackage{stmaryrd}
\usepackage{mathtools}

\usepackage{amsthm, color, xcolor, hyperref, slashed, subfigure, ytableau}
\ytableausetup{boxsize=0.8em}

\def \bal#1\eal  {\begin{align} #1 \end{align}}
\def \bga#1\ega  {\begin{gather} #1 \end{gather}}
\def\({\left(}
\def\){\right)}
\def\[{\left[}
\def\]{\right]}
\def\<{\left\langle}
\def\>{\right\rangle}

\newcommand{\bim} {\begin{itemize}[noitemsep]}

\newcommand{\eim}{\end{itemize}}
\newcommand{\beq} {\begin{equation}}
\newcommand{\eeq} {\end{equation}}
\newcommand{\bc}{\begin{center}}
\newcommand{\ec}{\end{center}}

\title{Unistochastic reduction of multi-matrix coherent-state kernels}

\author[a]{Shannon Wang}

\affiliation[a]{Department of Physics, University of California, Santa Barbara, CA 93106, USA}

\emailAdd{shannonwang@physics.ucsb.edu}

\abstract{
We study finite-$N$ coherent-state overlap kernels in multi-matrix BPS
sectors of $\mathcal N=4$ super Yang--Mills theory. For commuting
coherent-state parameters, the kernel is a Laplace transform of the
Haar-pushforward measure on the unistochastic set. After quotienting
row- and column-shift redundancies, the nontrivial source is an
$(N-1)\times(N-1)$ matrix $D$. On the physical $m$-matrix locus, $D$ is a
Gram matrix with $\operatorname{rank}D\leq m$, and the ordinary HCIZ sector
is precisely the rank-at-most-one stratum.

For $N=3$ we obtain an exact Bessel-integral representation of the first
genuinely rank-two kernel. Along the rank-one ray $D=rE_{11}$,
$F_N={}_1F_1(1;N;r)$ for arbitrary finite $N$. Using
$K_N=\log F_N$ as the reduced K\"ahler potential, we study a controlled
two-matrix embedding of the HCIZ sector and determine its transverse metric,
second fundamental form, and Gauss--Codazzi geometry. The embedding is
non-totally-geodesic at finite $N$. In the scaling $r=N\rho$, we find
\begin{equation}
Nh_N\to\min(\rho,1),
\qquad
\|B\|^2\to2(1-\rho)_+.
\end{equation}
}

\begin{document} 
\maketitle
\flushbottom

\section{Introduction}
\label{sec:first}

There has been a recent surge of interest in using coherent states as generating
functions for protected operators in $\mathcal N=4$ super Yang--Mills theory.
In the half-BPS sector, overlaps of gauge-averaged coherent states reduce to the
Harish-Chandra--Itzykson--Zuber (HCIZ) integral
\cite{harish1957differential,itzykson1980planar}. Its exact evaluation therefore
provides a finite-$N$ generating function for half-BPS two-point correlators in
collective coordinates \cite{Berenstein:2022srd}.
This naturally suggests analogous constructions in the $1/4$- and $1/8$-BPS
sectors, where several commuting complex matrices are active and the resulting
group integrals are no longer of ordinary HCIZ form.

This coherent-state framework has subsequently been developed in several
directions
\cite{Holguin:2022drf,Lin:2022wdr,Holguin:2022zii,Holguin:2023orq}.
The construction extends to $\operatorname{Sp}(2N)$,
$\operatorname{SO}(2N+1)$, and $\operatorname{SO}(2N)$ theories, where
half-BPS norms are again governed by Harish-Chandra-type group integrals
\cite{Holguin:2022drf}. Related coherent-state constructions have been used to
describe giant-graviton excitations \cite{Lin:2022wdr} and holographic
three-point functions \cite{Holguin:2022zii}, while matrix-model formulations
connect heavy half-BPS correlators to random-matrix ensembles and LLM
geometries \cite{Holguin:2023orq}. These developments indicate that group- and
matrix-integral descriptions are a robust feature of protected coherent-state
sectors.

A qualitatively different problem appears when the protected sector itself is
enlarged. In Ref.~\cite{holguin2024multi}, the coherent-state construction was
extended to finite-$N$ $1/4$-BPS operators. For two complex matrices, the norm
takes the form
\begin{equation}
\label{eqn:twopointfunction}
    \bra{\bar{\Lambda}_X,\bar{\Lambda}_Y}
    \ket{\Lambda_X,\Lambda_Y}
    =
    \frac{1}{\operatorname{Vol}[U(N)]}
    \int dU\,
    \exp\left(
    \operatorname{Tr}\left[
    U\bar{\Lambda}_XU^\dagger\Lambda_X
    +
    U\bar{\Lambda}_YU^\dagger\Lambda_Y
    \right]\right),
\end{equation}
for mutually commuting source matrices. The $U(2)$ case can be evaluated
exactly, while for $U(3)$ a direct parametrization of the unitary matrix leads
to a cumbersome expansion rather than a useful closed representation.
The first nontrivial finite-$N$ example therefore already exhibits structure
that is absent from the ordinary HCIZ problem.

The same class of integrals has recently appeared from a complementary
large-$N$ viewpoint. In Ref.~\cite{anempodistov2026solvable}, complex
matrix models for protected BPS correlators were related to quenched
Eguchi--Kawai reductions of the principal chiral model, with the
$1/4$- and $1/8$-BPS problems corresponding to two- and three-dimensional
reductions, respectively. This formulation emphasizes the large-$N$ dynamics
of the multi-matrix problem. The purpose of the present work is instead to
expose its finite-$N$ source geometry and to determine how the ordinary HCIZ
sector is embedded inside the larger multi-matrix coherent-state space.

For simultaneously diagonalizable coherent-state parameters, the exponent can
be written as
\begin{equation}
    \sum_{i,j=1}^{N} C_{ij}\abs{U_{ij}}^2 .
\end{equation}
Because $q_{ij}=\abs{U_{ij}}^2$ is doubly stochastic, the source matrix is
defined nontrivially only modulo row and column shifts,
\begin{equation}
    C_{ij}\sim C_{ij}+r_i+s_j.
\end{equation}
A convenient set of invariant coordinates is therefore
\begin{equation}
\label{eq:intro-reduced-source}
    D_{ij}
    =
    C_{ij}-C_{iN}-C_{Nj}+C_{NN},
    \qquad
    i,j=1,\ldots,N-1.
\end{equation}
For $m$ commuting matrices, the physical source takes the factorized form
\begin{equation}
\label{eq:intro-rank}
    D_{ij}
    =
    \sum_{A=1}^{m}
    \left(a_i^{(A)}-a_N^{(A)}\right)
    \left(\bar a_j^{(A)}-\bar a_N^{(A)}\right),
\end{equation}
and hence has rank at most $m$. On the physical conjugate locus this is a Gram
matrix. The ordinary HCIZ integral is precisely the rank-one case: after the
row- and column-shift quotient, a source admits an ordinary separable HCIZ
representative if and only if $\operatorname{rank}D\leq1$. Generic two-matrix
sources instead have rank two. This gives the multi-matrix coherent-state
problem a natural rank stratification, with the HCIZ sector forming a
lower-rank locus inside the larger protected state space.

The unistochastic formulation provides a natural finite-$N$ description of
this stratified source geometry. Since the unitary matrix enters only through
$q_{ij}=\abs{U_{ij}}^2$, the group integral is the Laplace transform of the
pushforward of Haar measure under
\begin{equation}
    \pi:U(N)\longrightarrow\mathcal U_N,
    \qquad
    U\longmapsto q,
    \qquad
    q_{ij}=\abs{U_{ij}}^2 .
\end{equation}
After removing the elementary row- and column-shift dependence, the reduced
kernel takes the form
\begin{equation}
    F_N(D)
    =
    \int_{\mathcal U_N}
    \exp\left(
    \sum_{i,j=1}^{N-1}D_{ij}q_{ij}
    \right)
    d\nu_N(q),
\end{equation}
where $d\nu_N$ is the Haar-pushforward measure. For $N=3$, this measure is
known explicitly, and we obtain a compact Bessel-integral representation of
the first genuinely rank-two kernel. The rank-one condition reduces in this
case to a single determinantal equation, so the exact $U(3)$ kernel provides a
controlled interpolation between the ordinary HCIZ locus and genuinely
two-matrix configurations.

The logarithm of the reduced kernel has an additional geometric
interpretation. Writing
\begin{equation}
    K_N(D)=\log F_N(D),
\end{equation}
its first and second source derivatives are respectively the mean and connected
covariance of the unistochastic variables in the exponentially tilted Haar
ensemble. Pulling $K_N$ back to coherent-state eigenvalue space therefore
defines the K\"ahler metric of the corresponding coherent-state family
\cite{provost1980riemannian}. The relation between quantum-state overlap
generating functions, gauge-invariant cumulants, and quantum geometry has
been developed more generally in
Refs.~\cite{SouzaWilkensMartin2000,HetenyiLevay2023}, the latter including
higher-order connection- and curvature-related tensors. Here we combine this
cumulant structure with the nonlinear Gram geometry of the multi-matrix
source to study the embedding between source-rank strata. This allows the
rank stratification to be studied as a problem in the quantum geometry of
protected states.

A particularly tractable family is the rank-one ray
\begin{equation}
    D=rE_{11}.
\end{equation}
For arbitrary finite $N$, the corresponding kernel is exactly
\begin{equation}
\label{eq:intro-hypergeom}
    F_N(rE_{11})
    =
    {}_1F_1(1;N;r).
\end{equation}
Using this result, we determine the metric normal to the HCIZ locus, its second
fundamental form, and the associated Gauss--Codazzi geometry inside a
controlled three-complex-dimensional slice of the two-matrix coherent-state
manifold. Within this slice, the rank-one locus is smooth for nonzero source
but is not totally geodesic at finite $N$: part of its intrinsic curvature
arises from its extrinsic bending toward the rank-two directions.

The same family also admits a controlled large-$N$ limit. With
$r=N\rho$, the integral develops distinct boundary- and interior-saddle
regimes separated at $\rho=1$. The corresponding normal geometry has the
simple limits
\begin{equation}
    N h_N(N\rho)\longrightarrow \min(\rho,1),
    \qquad
    \norm{B}^2\longrightarrow 2(1-\rho)_+ ,
\end{equation}
where $h_N$ is the transverse metric and $B$ is the second fundamental form
of the controlled HCIZ surface. Thus, within this controlled family, the
finite-$N$ HCIZ surface is extrinsically curved inside the two-matrix
coherent-state manifold, while in the aligned large-$N$ regime this controlled
embedding becomes asymptotically geodesic. The large-$N$ saddle itself is closely related to rank-one spherical-integral asymptotics \cite{guionnet2005fourier}; the new role played here by the saddle is to control the geometry of the embedding between protected
coherent-state sectors.

The paper is organized as follows. In section~\ref{sec:second}, we formulate
the coherent-state kernel as a unistochastic Laplace transform, derive the
source quotient, and characterize the rank stratification of the physical
source locus. In section~\ref{sec:third}, we study the first genuinely
rank-two case and derive the exact $U(3)$ Bessel representation. In
section~\ref{sec:fourth}, we develop the finite-$N$ theory at fixed source
rank, including the exact rank-one kernel and the relation to Haar moments and
sampling. We then study the K\"ahler geometry of the coherent-state family and
the finite-$N$ geometry of the HCIZ embedding, before analyzing its controlled
large-$N$ limit. We conclude with implications for the organization of
multi-matrix protected sectors and directions for extending the construction
to more general BPS and holographic observables.

\section{Unistochastic reduction and rank geometry of the source}
\label{sec:second}

\subsection{Unistochastic reduction}
\label{sec:unistochastic-reduction}

We begin with the coherent-state norm introduced in
Ref.~\cite{holguin2024multi},
\begin{equation}
\mathcal I_N =
\frac{1}{\operatorname{Vol}[U(N)]}
\int_{U(N)}dU\,
\exp\left[
\operatorname{Tr}
\left(
U\bar\Lambda_XU^\dagger\Lambda_X+
U\bar\Lambda_YU^\dagger\Lambda_Y
\right)
\right].
\label{eq:kernel}
\end{equation}
We take the mutually commuting coherent-state parameters to be diagonal,
\begin{equation}
\Lambda_X=\operatorname{diag}(a_1,\ldots,a_N),
\qquad
\Lambda_Y=\operatorname{diag}(b_1,\ldots,b_N),
\end{equation}
together with
\begin{equation}
\bar\Lambda_X=\operatorname{diag}(\bar a_1,\ldots,\bar a_N),
\qquad
\bar\Lambda_Y=\operatorname{diag}(\bar b_1,\ldots,\bar b_N).
\end{equation}
The barred variables are kept independent throughout the holomorphic
formulation and are restricted to complex conjugates only on the physical
slice corresponding to an actual norm.

Expanding the trace gives
\begin{equation}
\operatorname{Tr}\left(
U\bar\Lambda_XU^\dagger\Lambda_X+
U\bar\Lambda_YU^\dagger\Lambda_Y
\right)
=
\sum_{i,j=1}^N
\left(
a_i\bar a_j+b_i\bar b_j
\right)
|U_{ij}|^2 .
\label{eq:exponent-q}
\end{equation}
We therefore introduce
\begin{equation}
C_{ij}=a_i\bar a_j+b_i\bar b_j,
\label{eq:Cdef}
\end{equation}
and
\begin{equation}
q_{ij}=|U_{ij}|^2.
\label{eq:qdef}
\end{equation}
In these variables,
\begin{equation}
\mathcal I_N(C)
=
\int_{U(N)}d\mu_H(U)\,
\exp\left(
\sum_{i,j=1}^N C_{ij}q_{ij}(U)
\right),
\label{eq:kernel-Cq}
\end{equation}
where $d\mu_H$ denotes normalized Haar measure.

Unitarity implies
\begin{equation}
q_{ij}\geq0,
\qquad
\sum_{j=1}^N q_{ij}=1,
\qquad
\sum_{i=1}^N q_{ij}=1.
\label{eq:doubly-stochastic}
\end{equation}
Thus $q$ is doubly stochastic. The matrices that can arise as
$q_{ij}=|U_{ij}|^2$ form the unistochastic subset
\begin{equation}
\mathcal U_N
=
\left\{
q\in\mathcal B_N:
q_{ij}=|U_{ij}|^2
\ \text{for some }U\in U(N)
\right\}
\label{eq:unistochastic-set}
\end{equation}
of the Birkhoff polytope $\mathcal B_N$
\cite{zyczkowski2003random,bengtsson2005birkhoff}.

For $N=2$ every bistochastic matrix is unistochastic, while for $N=3$
the unistochastic set is a proper subset of the Birkhoff polytope whose
boundary is characterized by saturation of the unitarity-triangle
inequalities
\cite{bengtsson2005birkhoff,dunkl2009volume}. For $N\geq4$ the geometry of the unistochastic subset is substantially more
complicated \cite{bengtsson2005birkhoff,dictua2006separation}.

The map
\begin{equation}
\pi:U(N)\longrightarrow\mathcal U_N,
\qquad
\pi(U)_{ij}=|U_{ij}|^2
\label{eq:pi-map}
\end{equation}
pushes normalized Haar measure forward to the probability measure
\begin{equation}
\nu_N=\pi_*\mu_H .
\label{eq:pushforward-measure}
\end{equation}
This is the random unistochastic ensemble studied in
Ref.~\cite{zyczkowski2003random}. By definition of the pushforward,
Eq.~\eqref{eq:kernel-Cq} becomes
\begin{equation}
\mathcal I_N(C)
=
\int_{\mathcal U_N}
\exp\left(
\sum_{i,j=1}^N C_{ij}q_{ij}
\right)
d\nu_N(q).
\label{eq:unistochastic-laplace}
\end{equation}
The coherent-state kernel is therefore a Laplace transform of the
Haar-pushforward measure on the unistochastic set. The usefulness of this
formulation is not merely a reduction in integration variables: it separates
the universal Haar-induced measure from the geometry of the physical source
locus, which will play a central role below.

\subsection{Source quotient from double stochasticity}
\label{sec:source-reduction}

It is useful initially to regard Eq.~\eqref{eq:unistochastic-laplace} as
defining $\mathcal I_N(C)$ for a general complex $N\times N$ source $C$.
The physical coherent-state source will then be imposed after quotienting a
redundancy implied by double stochasticity.

Consider
\begin{equation}
C_{ij}\longrightarrow
C'_{ij}=C_{ij}+r_i+s_j ,
\label{eq:source-shift}
\end{equation}
where $r_i$ and $s_j$ are arbitrary. Using
Eq.~\eqref{eq:doubly-stochastic},
\begin{equation}
\sum_{i,j=1}^N C'_{ij}q_{ij}
=
\sum_{i,j=1}^N C_{ij}q_{ij}
+
\sum_{i=1}^N r_i
+
\sum_{j=1}^N s_j .
\label{eq:source-shift-exponent}
\end{equation}
Consequently,
\begin{equation}
\mathcal I_N(C+r\mathbf 1^T+\mathbf 1s^T)
=
\exp\left(
\sum_i r_i+\sum_j s_j
\right)
\mathcal I_N(C).
\label{eq:source-shift-kernel}
\end{equation}
Row and column potentials therefore change the kernel only by an elementary
overall exponential factor.

The space of such shifts has dimension $2N-1$, so the nontrivial source
dependence is carried by
\begin{equation}
N^2-(2N-1)=(N-1)^2
\end{equation}
independent combinations. Choosing the $N$th row and column as reference, a
convenient set is
\begin{equation}
D_{ij}
=
C_{ij}-C_{iN}-C_{Nj}+C_{NN},
\qquad
i,j=1,\ldots,N-1.
\label{eq:Ddef}
\end{equation}
The matrix $D$ is invariant under Eq.~\eqref{eq:source-shift} and provides
coordinates on the quotient of the source space by row and column
potentials.

Eliminating the last row and column of $q$ gives
\begin{equation}
\sum_{i,j=1}^N C_{ij}q_{ij}
=
R_N(C)
+
\sum_{i,j=1}^{N-1}D_{ij}q_{ij},
\label{eq:gauge-reduced-exponent}
\end{equation}
where
\begin{equation}
R_N(C)
=
\sum_{i=1}^{N-1}C_{iN}
+
\sum_{j=1}^{N-1}C_{Nj}
-
(N-2)C_{NN}.
\label{eq:Rdef}
\end{equation}
The kernel therefore factorizes as
\begin{equation}
\mathcal I_N(C)
=
e^{R_N(C)}F_N(D),
\label{eq:kernel-factorization}
\end{equation}
with
\begin{equation}
F_N(D)
=
\int_{\mathcal U_N}
\exp\left[
\sum_{i,j=1}^{N-1}D_{ij}q_{ij}
\right]
d\nu_N(q).
\label{eq:reduced-kernel}
\end{equation}
All nontrivial source dependence is thus encoded in the reduced kernel
$F_N(D)$.

\subsection{Rank geometry of physical coherent-state sources}
\label{sec:rank-geometry}

The reduction above becomes particularly restrictive when $D$ is returned to
the physical coherent-state locus. For the two-matrix source
Eq.~\eqref{eq:Cdef},
\begin{equation}
D_{ij}
=
(a_i-a_N)(\bar a_j-\bar a_N)
+
(b_i-b_N)(\bar b_j-\bar b_N).
\label{eq:D-coherent}
\end{equation}
Defining
\begin{equation}
\alpha_i=a_i-a_N,
\qquad
\beta_i=b_i-b_N,
\qquad
i=1,\ldots,N-1,
\end{equation}
we may write
\begin{equation}
D
=
\alpha\,\bar\alpha^{\,T}
+
\beta\,\bar\beta^{\,T}.
\label{eq:D-rank-two}
\end{equation}
In the holomorphic formulation the barred variables are independent, so this
is more generally a bilinear factorization. On the physical conjugate locus,
however,
\begin{equation}
D
=
\alpha\alpha^\dagger+\beta\beta^\dagger
=
VV^\dagger,
\qquad
V=(\alpha,\beta),
\label{eq:D-Gram}
\end{equation}
and $D$ is a positive-semidefinite Gram matrix. In either description,
\begin{equation}
\operatorname{rank}D\leq2.
\end{equation}

More generally, for $m$ mutually commuting complex matrices,
\begin{equation}
C_{ij}
=
\sum_{A=1}^{m}
a_i^{(A)}\bar a_j^{(A)}
\end{equation}
gives
\begin{equation}
D_{ij}
=
\sum_{A=1}^{m}
\left(a_i^{(A)}-a_N^{(A)}\right)
\left(\bar a_j^{(A)}-\bar a_N^{(A)}\right).
\label{eq:D-mmatrix}
\end{equation}
Hence
\begin{equation}
\operatorname{rank}D
\leq
\min(m,N-1).
\label{eq:rank-bound}
\end{equation}
The physical source space is therefore not a generic
$(N-1)^2$-dimensional matrix space. It lies on a hierarchy of low-rank
strata,
\begin{equation}
\mathcal R_r
=
\left\{
D:\operatorname{rank}D\leq r
\right\},
\qquad
\mathcal R_1\subset\mathcal R_2\subset\cdots .
\label{eq:rank-strata}
\end{equation}
On the physical conjugate locus, the rank of $D$ therefore counts the
number of linearly independent relative scalar directions probed by the
coherent-state source, and is bounded above by the number of active
commuting matrices.

\subsection{The HCIZ sector as the rank-one stratum}
\label{sec:hciz-rank-one}

The rank stratification also gives a precise characterization of the ordinary
HCIZ problem. An ordinary HCIZ source is separable \cite{harish1957differential,itzykson1980planar},
\begin{equation}
C_{ij}=x_i y_j .
\label{eq:HCIZ-source}
\end{equation}
Its reduced source is
\begin{equation}
D_{ij}
=
(x_i-x_N)(y_j-y_N),
\label{eq:HCIZ-reduced}
\end{equation}
and therefore
\begin{equation}
\operatorname{rank}D\leq1.
\end{equation}

The converse, which will be important below, follows directly from the
source quotient. Suppose the reduced source has rank one,
\begin{equation}
D_{ij}=u_i v_j .
\end{equation}
Choose
\begin{equation}
x=(u_1,\ldots,u_{N-1},0),
\qquad
y=(v_1,\ldots,v_{N-1},0).
\end{equation}
Then the separable source $C_{ij}=x_i y_j$ has precisely the reduced source
$D$. Any other representative of the same source class differs only by the
row- and column-shift redundancy of Eq.~\eqref{eq:source-shift}. Thus
\begin{equation}
[C]\ \text{contains an ordinary HCIZ representative}
\quad\Longleftrightarrow\quad
\operatorname{rank}D\leq1.
\label{eq:hciz-iff-rank-one}
\end{equation}

The ordinary HCIZ problem is therefore not merely a special choice of
parameters inside the multi-matrix kernel: modulo the row- and column-shift
redundancy, it is precisely the rank-at-most-one stratum of the reduced
source geometry.

For the physical two-matrix problem,
Eq.~\eqref{eq:D-Gram} implies that the rank is at most one precisely
when the two relative eigenvalue vectors $\alpha$ and $\beta$ are linearly
dependent. Thus the rank-one locus consists of configurations in which the two
nominal scalar directions collapse to a single effective complex direction,
whereas a generic two-matrix configuration is genuinely rank two. In this
sense the enlargement from the ordinary HCIZ sector to the two-matrix BPS
kernel is geometrically the passage
\begin{equation}
\mathcal R_1\longrightarrow\mathcal R_2.
\end{equation}

For $N=3$, this distinction is controlled by a single determinant. Writing
\begin{equation}
D=
\begin{pmatrix}
x&y\\
z&w
\end{pmatrix},
\end{equation}
the HCIZ locus is
\begin{equation}
xw-yz=0.
\label{eq:N3-rank-one}
\end{equation}
On the physical conjugate locus,
\begin{equation}
\det D
=
|\alpha_1\beta_2-\alpha_2\beta_1|^2.
\label{eq:N3-Gram-det}
\end{equation}
The determinant is therefore the Gram determinant measuring the departure from
an effectively one-scalar configuration. We will use the exact $N=3$ kernel
in the next section to study the first genuinely rank-two case.

\subsection{Cumulants of the tilted unistochastic ensemble}
\label{sec:cumulants}

The reduced kernel also carries more information than its Taylor coefficients
at the origin. Define
\begin{equation}
K_N(D)=\log F_N(D).
\label{eq:KN-def}
\end{equation}
For sources for which the exponential weight is real and positive, introduce
the tilted probability measure
\begin{equation}
d\nu_{N,D}(q)
=
\frac{
\exp\left(\sum_{ij}D_{ij}q_{ij}\right)
}{
F_N(D)
}
\,d\nu_N(q).
\label{eq:tilted-measure}
\end{equation}
For general complex sources the same formulas hold by analytic continuation.
Expectation values with respect to Eq.~\eqref{eq:tilted-measure} satisfy
\begin{equation}
\frac{\partial K_N}{\partial D_{ij}}
=
\left\langle q_{ij}\right\rangle_D,
\label{eq:KN-first-derivative}
\end{equation}
and
\begin{equation}
\frac{\partial^2K_N}
{\partial D_{ij}\partial D_{kl}}
=
\left\langle
q_{ij}q_{kl}
\right\rangle_D
-
\left\langle q_{ij}\right\rangle_D
\left\langle q_{kl}\right\rangle_D .
\label{eq:KN-second-derivative}
\end{equation}
Thus $K_N$ generates the connected cumulants of the Haar-induced
unistochastic ensemble.

At the origin,
\begin{equation}
F_N(0)=1,
\end{equation}
and Haar symmetry gives
\begin{equation}
\left.
\frac{\partial F_N}{\partial D_{ij}}
\right|_{D=0}
=
\left\langle q_{ij}\right\rangle_0
=
\frac{1}{N}.
\label{eq:first-moment}
\end{equation}
Higher derivatives of $K_N$ generate the corresponding higher connected
cumulants, while derivatives of $F_N$ at the origin reproduce ordinary Haar
moments and may be evaluated systematically using Weingarten calculus.

Generating functions constructed from quantum-state overlaps have more
generally been used to organize quantum geometry in terms of gauge-invariant
cumulants. An earlier generating-function formulation appears in the theory of
electronic polarization of Souza, Wilkens, and Martin, where the second
cumulant is related to the Brillouin-zone-integrated quantum metric
\cite{SouzaWilkensMartin2000}. Het\'enyi and L\'evay subsequently developed
the overlap-generating-function viewpoint directly for quantum-state
geometry, showing that second-order cumulants encode the complex quantum
metric while higher-order cumulants enter connection- and curvature-related
geometric tensors \cite{HetenyiLevay2023}. In the present setting, $K_N$ is the logarithm of the
reduced coherent-state norm and the physical coherent-state coordinates
enter through the nonlinear Gram map
$D_{ij}=\sum_A\xi_{iA}\bar\xi_{jA}$. Consequently, the pullback geometry below
combines source cumulants through the corresponding chain rules rather than
identifying its geometric tensors term by term with source-space cumulants.

Equations~\eqref{eq:KN-first-derivative} and
\eqref{eq:KN-second-derivative} will later acquire a direct geometric
interpretation. The full coherent-state norm has K\"ahler potential
\begin{equation}
\log \mathcal I_N(C)=R_N(C)+K_N(D).
\end{equation}
On the reduced relative-coordinate slice used below, obtained by fixing the
reference eigenvalues so that $R_N(C)=0$, the pullback of $K_N$ itself serves
as a K\"ahler potential for the corresponding family of protected coherent
states \cite{provost1980riemannian}. The unistochastic means and connected
fluctuations therefore determine the finite-$N$ quantum geometry on this
reduced coherent-state manifold.

Before turning to this geometry, we first exploit the explicit
Haar-pushforward measure available at $N=3$. This gives the first exact kernel
away from the rank-one HCIZ stratum and provides a finite-$N$ laboratory for
genuinely two-matrix source geometry.


\section{Exact kernels and the first genuinely rank-two geometry}
\label{sec:third}

\subsection{$U(2)$: collapse to the rank-one sector}
\label{sec:U2}

The $N=2$ case is special not only because the unistochastic set is
one-dimensional, but also because the reduced source space itself is
one-dimensional. Consequently,
\begin{equation}
\operatorname{rank}D\leq1
\end{equation}
identically, independently of how many commuting scalar matrices contribute
to the original coherent-state source. Thus $U(2)$ cannot support a
genuinely higher-rank reduced source: after quotienting the row- and
column-shift redundancy, every such kernel is HCIZ-equivalent in the sense
of Eq.~\eqref{eq:hciz-iff-rank-one}.

Every $2\times2$ doubly stochastic matrix is unistochastic and may be written
in terms of a single parameter $t\in[0,1]$ as
\begin{equation}
q(t)
=
\begin{pmatrix}
t & 1-t \\
1-t & t
\end{pmatrix}.
\label{eq:U2-q}
\end{equation}
For a Haar-distributed matrix $U\in U(2)$, the first row is uniformly
distributed on the unit sphere in $\mathbb C^2$, so that $t=|U_{11}|^2$ is uniformly distributed on the unit interval.
Consequently, the Haar-pushforward measure on $\mathcal U_2$ is simply
\begin{equation}
d\nu_2(t)=dt,
\qquad
0\leq t\leq1.
\label{eq:U2-measure}
\end{equation}

For $N=2$, the reduced source space introduced in
Eq.~\eqref{eq:Ddef} is one-dimensional. Defining
\begin{equation}
D
=
C_{11}-C_{12}-C_{21}+C_{22},
\label{eq:U2-D}
\end{equation}
Eq.~\eqref{eq:reduced-kernel} becomes
\begin{equation}
F_2(D)
=
\int_0^1 dt\,e^{Dt}
=
\frac{e^D-1}{D},
\label{eq:U2-reduced-kernel}
\end{equation}
with the regular limiting value $F_2(0)=1$.

The prefactor in Eq.~\eqref{eq:kernel-factorization} is determined by
\begin{equation}
R_2(C)=C_{12}+C_{21},
\label{eq:U2-R}
\end{equation}
and hence the full kernel is
\begin{equation}
\mathcal I_2(C)
=
e^{C_{12}+C_{21}}
\frac{
e^{C_{11}-C_{12}-C_{21}+C_{22}}-1
}{
C_{11}-C_{12}-C_{21}+C_{22}
}.
\label{eq:U2-kernel-1}
\end{equation}
Equivalently,
\begin{equation}
\mathcal I_2(C)
=
\frac{
e^{C_{11}+C_{22}}
-
e^{C_{12}+C_{21}}
}{
C_{11}-C_{12}-C_{21}+C_{22}
}.
\label{eq:U2-kernel}
\end{equation}

On the coherent-state source locus \eqref{eq:Cdef}, the single nontrivial source combination becomes
\begin{equation}
D
=
(a_1-a_2)(\bar a_1-\bar a_2)
+
(b_1-b_2)(\bar b_1-\bar b_2).
\label{eq:U2-D-coherent}
\end{equation}
Thus the unistochastic formulation reproduces the functional form of
the $U(2)$ result obtained in \cite{holguin2024multi}. The result in
\cite{holguin2024multi} carries an additional overall factor of
$1/4$, originating from the normalization of the Haar measure adopted
in the explicit $SU(2)$ parameterization. With the normalized Haar
probability measure used here, this factor is absent. The central
$U(1)$ phase drops out of $|U_{ij}|^2$, so this difference is purely
one of measure normalization rather than a distinction between the
$U(2)$ and $SU(2)$ integrals.

The reduced source is therefore a single scalar and contains no invariant
information capable of distinguishing one scalar direction from several
aligned scalar directions. The simplicity of the $U(2)$ multi-matrix kernel
is thus partly geometric: its reduced source space coincides with the
rank-one HCIZ stratum.

The situation changes qualitatively at $N=3$. The reduced source is then a
$2\times2$ matrix and may have rank two, so $N=3$ is the first value of
$N$ at which the coherent-state kernel can distinguish genuinely
multi-matrix source geometry. At the same time, the unistochastic set becomes
a proper subset of the Birkhoff polytope, while remaining sufficiently
explicit to permit an exact evaluation of the kernel.

\subsection{The $3\times3$ unistochastic set}
\label{sec:U3-geometry}

The $N=3$ case is the first for which double stochasticity is not
sufficient to guarantee unistochasticity. A $3\times3$ doubly
stochastic matrix has four independent entries, and may be written as
\begin{equation}
q=
\begin{pmatrix}
b_1 & b_2 & 1-b_1-b_2 \\
b_3 & b_4 & 1-b_3-b_4 \\
1-b_1-b_3 & 1-b_2-b_4 &
b_1+b_2+b_3+b_4-1
\end{pmatrix}.
\label{eq:U3-bistochastic}
\end{equation}
The positivity of all nine entries defines the Birkhoff polytope
$\mathcal B_3$, but only a proper subset of these matrices can be
written as $q_{ij}=|U_{ij}|^2$ for a unitary matrix $U$.

The additional constraint follows from the orthogonality of the
columns of $U$. For example,
\begin{equation}
\sum_{i=1}^3 U_{i1}^*U_{i2}=0
\label{eq:column-orthogonality}
\end{equation}
requires three complex numbers of magnitudes
\begin{equation}
L_i=\sqrt{q_{i1}q_{i2}},
\qquad i=1,2,3,
\label{eq:triangle-lengths}
\end{equation}
to sum to zero. Hence the three $L_i$ must form the sides of a
triangle. For $3\times3$ bistochastic matrices this triangle condition
is both necessary and sufficient for unistochasticity
\cite{bengtsson2005birkhoff,dunkl2009volume}.

Equivalently, the three lengths must satisfy the triangle inequalities,
which may be written, for example, as
\begin{equation}
|L_2-L_3|
\leq L_1
\leq L_2+L_3.
\label{eq:U3-triangle-inequality}
\end{equation}
A convenient way to encode this condition is through the area $A$ of
the corresponding unitarity triangle. By Heron's formula,
\begin{equation}
A^2
=
p(p-L_1)(p-L_2)(p-L_3),
\qquad
p=\frac{1}{2}(L_1+L_2+L_3).
\label{eq:U3-Heron}
\end{equation}
For the parameterization \eqref{eq:U3-bistochastic}, it is useful to
introduce the quantity
\begin{equation}
\begin{split}
Q(b_1,b_2,b_3,b_4)
&=
4b_1b_2b_3b_4
\\
&\quad
-
\left(
b_1+b_2+b_3+b_4-1-b_1b_4-b_2b_3
\right)^2 ,
\end{split}
\label{eq:U3-Q}
\end{equation}
which satisfies
\begin{equation}
Q=16A^2.
\label{eq:Q-area}
\end{equation}
Thus, within the Birkhoff polytope,
\begin{equation}
q\in\mathcal U_3
\qquad\Longleftrightarrow\qquad
Q(q)\geq0.
\label{eq:U3-unistochastic-condition}
\end{equation}
The boundary $Q=0$ corresponds to degenerate unitarity triangles and
is precisely the orthostochastic boundary of $\mathcal U_3$
\cite{bengtsson2005birkhoff,dunkl2009volume}.

The same quantity is related to the Jarlskog invariant of the unitary
lift. With
\begin{equation}
J
=
\operatorname{Im}
\left(
U_{11}U_{22}U_{12}^*U_{21}^*
\right),
\end{equation}
one has
\begin{equation}
J^2=\frac{Q}{4}=4A^2.
\label{eq:Jarlskog-Q}
\end{equation}
For a generic point in the interior of $\mathcal U_3$, fixing the moduli
determines two dephased unitary lifts, related by complex conjugation and
distinguished by the two signs of $J$. These branches coalesce on the
orthostochastic boundary $Q=0$
\cite{bengtsson2005birkhoff,dunkl2009volume}.

It is useful to distinguish the geometry described by $Q$ from the source
rank geometry introduced in Section~\ref{sec:rank-geometry}. The condition
$Q(q)\geq0$ characterizes the \emph{integration domain}: it determines which
doubly stochastic matrices admit unitary lifts. By contrast,
$\operatorname{rank}D$ characterizes the \emph{source geometry}. At $N=3$,
$q$ ranges over a four-real-dimensional unistochastic domain, while $D$ is a
general $2\times2$ complex matrix in the holomorphic formulation and restricts
to a four-real-dimensional positive-semidefinite Hermitian cone on the
physical conjugate locus. These structures play logically distinct roles.
The exact kernel below is the Laplace transform that couples the
unistochastic integration geometry to the reduced source geometry.

\subsection{The Haar pushforward and the exact $U(3)$ kernel}
\label{sec:U3-kernel}

The special tractability of the $N=3$ problem follows not only from
the criterion \eqref{eq:U3-unistochastic-condition}, but also from the
fact that the Haar-induced measure on $\mathcal U_3$ is known explicitly \cite{dunkl2009volume}. For any continuous function $g$ of the four independent
entries,
\begin{equation}
\begin{split}
&\int_{U(3)}
g\left(
|U_{11}|^2,|U_{12}|^2,
|U_{21}|^2,|U_{22}|^2
\right)
d\mu_H(U)
\\
&\qquad =
\frac{2}{\pi}
\int_{\Omega}
g(b_1,b_2,b_3,b_4)
\frac{
db_1\,db_2\,db_3\,db_4
}{
\sqrt{Q(b_1,b_2,b_3,b_4)}
},
\end{split}
\label{eq:U3-Haar-pushforward}
\end{equation}
where
\begin{equation}
\Omega
=
\left\{
(b_1,b_2,b_3,b_4):
q(b_1,b_2,b_3,b_4)\in\mathcal B_3,
\quad Q\geq0
\right\}.
\label{eq:U3-Omega}
\end{equation}
Thus the density induced by normalized Haar measure diverges as
$Q^{-1/2}$ toward the orthostochastic boundary, while remaining
integrable \cite{dunkl2009volume}.

Following Dunkl and \.{Z}yczkowski \cite{dunkl2009volume}, a
particularly useful parameterization of $\Omega$ is
\begin{equation}
b_1=b,
\qquad
b_2=s(1-b),
\qquad
b_3=t(1-b),
\label{eq:U3-bst}
\end{equation}
and
\begin{equation}
b_4
=
(1-s)(1-t)+bst
+
2\tau\sqrt{bst(1-s)(1-t)},
\label{eq:U3-b4}
\end{equation}
with
\begin{equation}
0\leq b,s,t\leq1,
\qquad
-1\leq \tau\leq1.
\label{eq:U3-bstr-domain}
\end{equation}
In these coordinates,
\begin{equation}
Q
=
4b(1-b)^2s(1-s)t(1-t)(1-\tau^2),
\label{eq:U3-Q-bstr}
\end{equation}
while the Jacobian is
\begin{equation}
\left|
\frac{
\partial(b_1,b_2,b_3,b_4)
}{
\partial(b,s,t,\tau)
}
\right|
=
2(1-b)^2
\sqrt{
bs(1-s)t(1-t)
}.
\label{eq:U3-Jacobian}
\end{equation}

Combining Eqs.~\eqref{eq:U3-Q-bstr} and
\eqref{eq:U3-Jacobian}, the Haar pushforward simplifies to
\begin{equation}
d\nu_3
=
\frac{2}{\pi}
\frac{1-b}{\sqrt{1-\tau^2}}\,
db\,ds\,dt\,d\tau.
\label{eq:U3-simple-measure}
\end{equation}
The normalization is immediate:
\begin{equation}
\frac{2}{\pi}
\int_0^1 db\,(1-b)
\int_0^1ds
\int_0^1dt
\int_{-1}^{1}\frac{d\tau}{\sqrt{1-\tau^2}}
=1.
\label{eq:U3-measure-normalization}
\end{equation}

For notational convenience, we write the four reduced source
variables as
\begin{equation}
x=D_{11},
\qquad
y=D_{12},
\qquad
z=D_{21},
\qquad
w=D_{22}.
\label{eq:xyzw}
\end{equation}

Equation~\eqref{eq:reduced-kernel} then becomes
\begin{equation}
\begin{split}
F_3(x,y,z,w)
&=
\frac{2}{\pi}
\int_0^1db
\int_0^1ds
\int_0^1dt
\int_{-1}^{1}
\frac{d\tau}{\sqrt{1-\tau^2}}\,(1-b)
\\
&\quad\times
\exp\Big[
xb
+ys(1-b)
+zt(1-b)
\\
&\qquad\qquad
+w\big(
(1-s)(1-t)+bst
+2\tau\sqrt{bst(1-s)(1-t)}
\big)
\Big].
\end{split}
\label{eq:U3-fourfold}
\end{equation}

The remaining dependence on $\tau$ is elementary. Using the integral
representation
\begin{equation}
I_0(\xi)
=
\frac{1}{\pi}
\int_{-1}^{1}
\frac{e^{\xi \tau}}{\sqrt{1-\tau^2}}\,d\tau
\label{eq:I0-representation}
\end{equation}
of the modified Bessel function, the $\tau$ integral can be performed
exactly. We obtain
\begin{equation}
\begin{split}
F_3(x,y,z,w)
&=
2
\int_0^1db
\int_0^1ds
\int_0^1dt\,
(1-b)
\\
&\quad\times
\exp\Big[
xb
+ys(1-b)
+zt(1-b)
\\
&\qquad\qquad
+w\big((1-s)(1-t)+bst\big)
\Big]
\\
&\quad\times
I_0\left(
2w\sqrt{bst(1-s)(1-t)}
\right).
\end{split}
\label{eq:U3-Bessel}
\end{equation}

The full coherent-state kernel follows from
Eq.~\eqref{eq:kernel-factorization}:
\begin{equation}
\mathcal I_3(C)
=
e^{R_3(C)}F_3(x,y,z,w),
\label{eq:U3-full-kernel}
\end{equation}
where
\begin{equation}
R_3(C)
=
C_{13}+C_{23}+C_{31}+C_{32}-C_{33},
\label{eq:U3-R}
\end{equation}
and
\begin{equation}
\begin{aligned}
x&=C_{11}-C_{13}-C_{31}+C_{33},\\
y&=C_{12}-C_{13}-C_{32}+C_{33},\\
z&=C_{21}-C_{23}-C_{31}+C_{33},\\
w&=C_{22}-C_{23}-C_{32}+C_{33}.
\end{aligned}
\label{eq:U3-reduced-sources}
\end{equation}

Equation~\eqref{eq:U3-Bessel} is therefore an exact representation of the
first coherent-state kernel whose reduced source can have rank greater than
one. Unlike the $U(2)$ result, it probes not only the HCIZ stratum
$\mathcal R_1$ but also the genuinely two-matrix region
$\mathcal R_2\setminus\mathcal R_1$. The Bessel representation will allow us
to study both regions within a single finite-$N$ expression.

The equivalence with the direct Euler-angle calculation of
Ref.~\cite{holguin2024multi} is reviewed in
Appendix~\ref{app:direct-U3}. In the unistochastic formulation the redundant
phase variables are integrated out at the level of the Haar pushforward,
leaving the source-dependent integral directly in variables adapted to the
kernel.

\subsection{Physical source geometry at $N=3$}
\label{sec:U3-source-geometry}

For $N=3$, the four variables appearing in the exact kernel
Eq.~\eqref{eq:U3-Bessel} are precisely the entries of the reduced source,
\begin{equation}
D=
\begin{pmatrix}
x&y\\
z&w
\end{pmatrix}.
\label{eq:U3-D-matrix}
\end{equation}
On the coherent-state locus, Eq.~\eqref{eq:D-coherent} specializes to
\begin{equation}
D
=
\alpha\,\bar\alpha^{\,T}
+
\beta\,\bar\beta^{\,T},
\end{equation}
with
\begin{equation}
\alpha=
\begin{pmatrix}
a_1-a_3\\
a_2-a_3
\end{pmatrix},
\qquad
\beta=
\begin{pmatrix}
b_1-b_3\\
b_2-b_3
\end{pmatrix}.
\end{equation}
Thus the nontrivial part of the exact $N=3$ coherent-state norm is
determined entirely by the relative eigenvalue data through
Eq.~\eqref{eq:U3-Bessel}; the remaining dependence is contained in the
elementary prefactor $e^{R_3(C)}$.

The two possible nonzero rank strata are distinguished by
\begin{equation}
\det D=xw-yz.
\label{eq:U3-source-determinant}
\end{equation}
The rank-one HCIZ locus is therefore
\begin{equation}
xw-yz=0,
\end{equation}
while a generic complex source with $xw-yz\neq0$ has rank two.

On the physical conjugate locus,
\begin{equation}
D
=
\alpha\alpha^\dagger+\beta\beta^\dagger,
\label{eq:U3-D-Gram}
\end{equation}
so that
\begin{equation}
D=D^\dagger,
\qquad
D\succeq0.
\end{equation}
Conversely, every positive-semidefinite $2\times2$ matrix admits such a
factorization. The physical two-matrix source space at $N=3$ therefore fills
the cone
\begin{equation}
\operatorname{Herm}_2^+
=
\left\{
D=D^\dagger:\ D\succeq0
\right\}.
\end{equation}
The apex $D=0$ is the rank-zero configuration; away from the apex, the
boundary of this cone has rank one, while its interior has rank two.

The determinant takes the particularly transparent form
\begin{equation}
\begin{split}
\det D
&=
\|\alpha\|^2\|\beta\|^2
-
|\alpha^\dagger\beta|^2
\\
&=
\left|
\alpha_1\beta_2-\alpha_2\beta_1
\right|^2.
\end{split}
\label{eq:U3-Gram-determinant}
\end{equation}
It is therefore the Gram determinant, or equivalently the squared complex
area spanned by the two relative eigenvalue vectors. Accordingly, away from
the apex, $\det D=0$ describes the physical rank-one boundary, while
\begin{equation}
\det D>0
\end{equation}
describes the genuinely rank-two interior.

The rank-one condition is equivalent to linear dependence of the two
relative scalar configurations $\alpha$ and $\beta$. When $\alpha\neq0$
this may be written
\begin{equation}
\beta=\lambda\alpha
\label{eq:aligned-scalars}
\end{equation}
for some $\lambda\in\mathbb C$; the cases in which one of the vectors
vanishes are included separately. For $\alpha\neq0$, equivalently,
\begin{equation}
b_i-b_3=\lambda(a_i-a_3),
\qquad i=1,2.
\end{equation}
Thus the three eigenvalue points $(a_i,b_i)\in\mathbb C^2$ lie on a single
affine complex line. Although both scalar matrices may be nonzero, their
relative data contain only one independent scalar direction, and the source
is HCIZ-equivalent. A generic two-matrix configuration fails this alignment
condition and lies in the rank-two interior.

\subsection{HCIZ restriction of the exact $U(3)$ kernel}
\label{sec:U3-HCIZ-boundary}

The characterization in Eq.~\eqref{eq:hciz-iff-rank-one} implies that the
restriction of the exact $U(3)$ kernel to $\det D=0$ must reduce to the
ordinary HCIZ integral. Let
\begin{equation}
D=uv^T,
\qquad
u=
\begin{pmatrix}u_1\\u_2\end{pmatrix},
\qquad
v=
\begin{pmatrix}v_1\\v_2\end{pmatrix},
\end{equation}
and introduce the augmented spectra
\begin{equation}
\widehat u=(u_1,u_2,0),
\qquad
\widehat v=(v_1,v_2,0).
\end{equation}
The corresponding separable $3\times3$ source
\begin{equation}
C_{ij}=\widehat u_i\widehat v_j
\end{equation}
has vanishing elementary prefactor $R_3(C)$ and reduced source precisely
equal to $D$. Hence the restriction of the Bessel representation
\eqref{eq:U3-Bessel} to the rank-one locus is exactly the ordinary $U(3)$
HCIZ integral, namely the $N=3$ specialization of the general rank-one
formula derived in Section~\ref{sec:rank-one-general-N}.

When
\begin{equation}
x=u_1v_1,\qquad
y=u_1v_2,\qquad
z=u_2v_1,\qquad
w=u_2v_2,
\end{equation}
so that $xw-yz=0$, the three-dimensional Bessel integral is exactly equal to
the $N=3$ specialization of the HCIZ determinant in
Eq.~\eqref{eq:rank-one-HCIZ-general-N}. For $xw-yz\neq0$, no row- or
column-shift of the bare source can make it separable: since
$\operatorname{rank}D$ is invariant under Eq.~\eqref{eq:source-shift}, the
source class contains no ordinary HCIZ representative.

This statement concerns the bare exponential kernel. Confluent limits and
derivatives of HCIZ integrals may generate additional insertions, but they do
not change the rank-one characterization of an ordinary separable source.

\subsection{Checks and the rank-one ray}
\label{sec:U3-checks}

Several checks of Eq.~\eqref{eq:U3-Bessel} follow immediately.
First, at vanishing source the normalization of the Haar-pushforward
measure gives
\begin{equation}
F_3(0,0,0,0)=1.
\label{eq:U3-zero-check}
\end{equation}
Likewise, differentiating at the origin gives
\begin{equation}
\left.
\frac{\partial F_3}{\partial x}
\right|_0
=
\left.
\frac{\partial F_3}{\partial y}
\right|_0
=
\left.
\frac{\partial F_3}{\partial z}
\right|_0
=
\left.
\frac{\partial F_3}{\partial w}
\right|_0
=
\frac13,
\label{eq:U3-first-moment-check}
\end{equation}
in agreement with the Haar expectation
$\langle |U_{ij}|^2\rangle=1/3$.

A particularly useful specialization is the rank-one ray
\begin{equation}
D=rE_{11}.
\end{equation}
Equation~\eqref{eq:U3-Bessel} then reduces to
\begin{equation}
\begin{split}
F_3(rE_{11})
&=
2\int_0^1db\,(1-b)e^{rb}
\\
&=
\frac{2(e^r-1-r)}{r^2}
\\
&=
{}_1F_1(1;3;r).
\end{split}
\label{eq:U3-rank-one-ray}
\end{equation}
The regular value at $r=0$ is $F_3(0)=1$. Equivalently, this is the
Laplace transform of the Haar marginal distribution of
$q_{11}=|U_{11}|^2$.

This specialization is more than an elementary consistency check. It is the
$N=3$ member of an exact rank-one family valid for arbitrary finite $N$,
derived in Section~\ref{sec:fourth},
\begin{equation}
F_N(rE_{11})={}_1F_1(1;N;r).
\end{equation}
The exact $U(3)$ kernel therefore connects the first genuinely rank-two
solution to an analytic rank-one sector that persists for all $N$.

For generic higher-rank sources, an equally explicit Haar-pushforward density
is not presently available for $N\geq4$. Nevertheless,
Eq.~\eqref{eq:reduced-kernel} remains exact: local information may be
obtained from Haar moments through Weingarten calculus, while finite-source
kernels may be accessed directly by Haar sampling. The rank-one sector,
however, admits additional analytic control at arbitrary finite $N$. In the
next section we therefore organize the general finite-$N$ problem by source
rank, beginning with the exact family previewed above.

\section{General finite $N$ at fixed source rank}
\label{sec:fourth}

\subsection{The exact rank-one sector at arbitrary $N$}
\label{sec:rank-one-general-N}

Section~\ref{sec:hciz-rank-one} showed that every reduced source of rank at
most one is HCIZ-equivalent. Thus, although the generic higher-rank
multi-matrix kernel is not known in closed form for arbitrary $N$, the
rank-one sector remains analytically tractable at every finite $N$.

Let
\begin{equation}
D=uv^T,
\qquad
u,v\in\mathbb C^{N-1},
\label{eq:rank-one-factorization-general}
\end{equation}
and introduce the augmented spectra
\begin{equation}
\widehat u=(u_1,\ldots,u_{N-1},0),
\qquad
\widehat v=(v_1,\ldots,v_{N-1},0).
\label{eq:augmented-spectra}
\end{equation}
The corresponding separable source
\begin{equation}
C_{ij}=\widehat u_i\widehat v_j
\end{equation}
has reduced source $D$ and vanishing elementary prefactor $R_N(C)$.
The ordinary HCIZ formula therefore gives
\begin{equation}
F_N(uv^T)
=
\left(\prod_{p=1}^{N-1}p!\right)
\frac{
\det\!\left(e^{\widehat u_i\widehat v_j}\right)_{i,j=1}^{N}
}{
\Delta(\widehat u)\Delta(\widehat v)
}
\label{eq:rank-one-HCIZ-general-N}
\end{equation}
where
\begin{equation}
\Delta(x)
=
\prod_{1\leq i<j\leq N}(x_i-x_j)
\end{equation}
denotes the Vandermonde determinant. Degenerate spectra are understood by
the usual confluent limits
\cite{harish1957differential,itzykson1980planar}.

A particularly useful one-parameter family is the rank-one ray
\begin{equation}
D=rE_{11}.
\label{eq:rank-one-ray-general}
\end{equation}
In this case the reduced kernel depends only on
$q_{11}=|U_{11}|^2$. For Haar-distributed $U\in U(N)$, the squared moduli of any row are
Dirichlet distributed
\cite{zyczkowski2001induced,mergny2022rank}, and hence
\begin{equation}
q_{11}\sim\operatorname{Beta}(1,N-1),
\end{equation}
with density
\begin{equation}
p_N(q)
=
(N-1)(1-q)^{N-2},
\qquad
0\leq q\leq1.
\label{eq:q11-beta-density}
\end{equation}
Equation~\eqref{eq:reduced-kernel} therefore reduces to
\begin{equation}
\begin{split}
F_N(rE_{11})
&=
(N-1)\int_0^1dq\,
e^{rq}(1-q)^{N-2}
\\
&=
{}_1F_1(1;N;r).
\end{split}
\label{eq:rank-one-hypergeometric}
\end{equation}
This is the one-projector specialization of the standard rank-one HCIZ
Dirichlet representation \cite{mergny2022rank}. For integer $N\geq2$, this may equivalently be written as
\begin{equation}
F_N(rE_{11})
=
\frac{(N-1)!}{r^{N-1}}
\left(
e^r-\sum_{k=0}^{N-2}\frac{r^k}{k!}
\right),
\label{eq:rank-one-elementary}
\end{equation}
with the regular limiting value $F_N(0)=1$.

For $N=3$, Eq.~\eqref{eq:rank-one-elementary} reduces to
Eq.~\eqref{eq:U3-rank-one-ray}. Thus the one-source specialization of the
exact $U(3)$ kernel is the first member of an exact finite-$N$ hierarchy.
The closed form \eqref{eq:rank-one-hypergeometric} will provide the analytic
input for the finite-$N$ K\"ahler and extrinsic geometry developed below.

\subsection{Exact moments from Weingarten calculus}
\label{sec:weingarten}

The rank-one sector is exactly solvable at arbitrary finite $N$, but a
generic source of rank two or higher is not known in closed form. Local
information about such sources is nevertheless available systematically from
the moment expansion of the exact pushforward representation
\eqref{eq:reduced-kernel}. Its Taylor coefficients are Haar moments of
products of squared unitary matrix elements and can therefore be evaluated
using unitary Weingarten calculus.

Expanding the exponential in Eq.~\eqref{eq:reduced-kernel} gives
\begin{equation}
F_N(D)
=
\sum_{k=0}^{\infty}
\frac{1}{k!}
\sum_{i_1,j_1=1}^{N-1}\cdots
\sum_{i_k,j_k=1}^{N-1}
D_{i_1j_1}\cdots D_{i_kj_k}
\left\langle
q_{i_1j_1}\cdots q_{i_kj_k}
\right\rangle_H ,
\label{eq:FN-moment-expansion}
\end{equation}
where
\begin{equation}
\left\langle
q_{i_1j_1}\cdots q_{i_kj_k}
\right\rangle_H
=
\int_{U(N)}
\prod_{\alpha=1}^{k}
|U_{i_\alpha j_\alpha}|^2\,
d\mu_H(U).
\label{eq:q-Haar-moments}
\end{equation}
These moments are standard unitary Haar integrals and may be
evaluated using Weingarten calculus
\cite{collins2006integration,collins2021weingarten}. In its general form,
\begin{equation}
\begin{split}
&\int_{U(N)}
\prod_{\alpha=1}^{k}
U_{i_\alpha j_\alpha}
\overline{U}_{i'_\alpha j'_\alpha}\,
d\mu_H(U)
\\
&\qquad =
\sum_{\sigma,\tau\in S_k}
\left(
\prod_{\alpha=1}^{k}
\delta_{i_\alpha,i'_{\sigma(\alpha)}}
\right)
\left(
\prod_{\alpha=1}^{k}
\delta_{j_\alpha,j'_{\tau(\alpha)}}
\right)
\operatorname{Wg}_N(\sigma^{-1}\tau),
\end{split}
\label{eq:weingarten-general}
\end{equation}
where $\operatorname{Wg}_N$ denotes the unitary Weingarten function.
Setting $i'_\alpha=i_\alpha$ and $j'_\alpha=j_\alpha$ gives precisely the
moments appearing in Eq.~\eqref{eq:q-Haar-moments}.

At first order,
\begin{equation}
\langle q_{ij}\rangle_H=\frac{1}{N}.
\label{eq:first-Haar-moment-general}
\end{equation}
At second order,
\begin{equation}
\left\langle q_{ij}q_{kl}\right\rangle_H
=
\frac{1+\delta_{ik}\delta_{jl}}{N^2-1}
-
\frac{\delta_{ik}+\delta_{jl}}
{N(N^2-1)}.
\label{eq:second-Haar-moment-general}
\end{equation}
In particular,
\begin{equation}
\langle q_{ij}^2\rangle_H
=
\frac{2}{N(N+1)},
\end{equation}
while for distinct entries in the same row or column,
\begin{equation}
\langle q_{ij}q_{il}\rangle_H
=
\langle q_{ij}q_{kj}\rangle_H
=
\frac{1}{N(N+1)},
\end{equation}
and for $i\neq k$ and $j\neq l$,
\begin{equation}
\langle q_{ij}q_{kl}\rangle_H
=
\frac{1}{N^2-1}.
\end{equation}

It is useful to define
\begin{equation}
S(D)=\sum_{i,j=1}^{N-1}D_{ij},
\qquad
\rho_i(D)=\sum_{j=1}^{N-1}D_{ij},
\qquad
\kappa_j(D)=\sum_{i=1}^{N-1}D_{ij}.
\label{eq:source-sums}
\end{equation}
Using Eqs.~\eqref{eq:first-Haar-moment-general} and
\eqref{eq:second-Haar-moment-general}, the reduced kernel has the expansion
\begin{equation}
\begin{split}
F_N(D)
={}&
1+\frac{1}{N}S(D)
\\
&+
\frac{1}{2(N^2-1)}
\left[
S(D)^2
+\sum_{i,j=1}^{N-1}D_{ij}^2
\right.
\\
&\hspace{2.8cm}\left.
-\frac{1}{N}
\left(
\sum_{i=1}^{N-1}\rho_i(D)^2
+
\sum_{j=1}^{N-1}\kappa_j(D)^2
\right)
\right]
+O(D^3).
\end{split}
\label{eq:FN-quadratic}
\end{equation}

For $N=3$, writing
$D_{11}=x$, $D_{12}=y$, $D_{21}=z$, and $D_{22}=w$, this gives
\begin{equation}
\begin{split}
F_3(x,y,z,w)
={}&
1+\frac{x+y+z+w}{3}
\\
&+
\frac{x^2+y^2+z^2+w^2}{12}
+
\frac{xy+xz+yw+zw}{12}
\\
&+
\frac{xw+yz}{8}
+O(D^3),
\end{split}
\label{eq:F3-quadratic}
\end{equation}
which agrees with the expansion of the exact Bessel representation
\eqref{eq:U3-Bessel}.

Higher-order coefficients may be obtained in the same way from
higher-order Weingarten functions. The number of permutations entering the
calculation grows rapidly with the order, so the moment expansion is most
useful as an exact local expansion around $D=0$ and as a check on
complementary numerical methods rather than as a global representation of
the kernel. Because the expansion is valid for arbitrary $D$, it may be
restricted directly to the low-rank physical loci identified in
Section~\ref{sec:rank-geometry}; source rank changes the combinations of
moments that enter, but not the Haar measure itself.

\subsection{Direct Haar sampling of rank-two sources}
\label{sec:Haar-sampling}

The pushforward formulation also provides a direct numerical method for
arbitrary finite $N$ without requiring an explicit formula for $d\nu_N$.
Sampling a Haar-distributed unitary matrix $U\in U(N)$ and forming
$q_{ij}=|U_{ij}|^2$ produces a sample distributed according to $\nu_N$.
Equation~\eqref{eq:reduced-kernel} may therefore be written as
\begin{equation}
F_N(D)
=
\mathbb E_{U\sim\mu_H}
\left[
\exp\left(
\sum_{i,j=1}^{N-1}
D_{ij}|U_{ij}|^2
\right)
\right].
\label{eq:FN-Haar-expectation}
\end{equation}

Given $M$ independent Haar-distributed unitary matrices
$U^{(1)},\ldots,U^{(M)}$, an unbiased Monte Carlo estimator is
\begin{equation}
\widehat F_{N,M}(D)
=
\frac{1}{M}
\sum_{m=1}^M
\exp\left[
\sum_{i,j=1}^{N-1}
D_{ij}
\left|U^{(m)}_{ij}\right|^2
\right].
\label{eq:MC-estimator}
\end{equation}
For fixed source $D$, the statistical uncertainty decreases with the usual
$M^{-1/2}$ scaling. Haar-distributed unitary matrices may be generated
numerically by QR decomposition of complex Ginibre matrices, with the phases
of the diagonal entries of the triangular factor removed in the standard
way \cite{mezzadri2006generate}.

We first test Eq.~\eqref{eq:MC-estimator} against the exact $N=3$ result. For
representative real source matrices $D$, direct Haar sampling agrees with the
Bessel representation \eqref{eq:U3-Bessel} within Monte Carlo uncertainty.
Near the origin, both agree in turn with the exact Weingarten expansion
\eqref{eq:F3-quadratic}. For $N>3$, the same estimator requires no
modification and provides direct access to finite-source kernels even when an
explicit pushforward density is unavailable.

For the numerical illustration below we restrict to real sources and consider
the one-parameter family
\begin{equation}
D(t)
=
t
\begin{pmatrix}
1 & \frac12 \\
\frac12 & 1
\end{pmatrix}
\oplus\mathbf 0_{N-3},
\label{eq:numerical-source}
\end{equation}
where the zero block is absent for $N=3$. For $t\geq0$ this source is
positive semidefinite and lies on the physical two-matrix locus. For $t<0$,
its nonzero $2\times2$ block is negative definite, while the full source is
negative semidefinite of rank two for $N>3$. The $t<0$ branch therefore lies
off the physical conjugate locus, although it remains rank two in the
holomorphic formulation. For $t>0$,
\begin{equation}
\det
\left[
t
\begin{pmatrix}
1 & \frac12 \\
\frac12 & 1
\end{pmatrix}
\right]
=
\frac34t^2>0,
\label{eq:numerical-source-determinant}
\end{equation}
so the trajectory lies strictly inside the genuinely rank-two region rather
than on the rank-one HCIZ locus.

Along this trajectory,
\begin{equation}
S(D)=3t,
\qquad
\sum_{i,j}D_{ij}^2=\frac52t^2,
\qquad
\sum_i\rho_i^2+\sum_j\kappa_j^2=9t^2.
\end{equation}
Equation~\eqref{eq:FN-quadratic} consequently reduces to
\begin{equation}
F_N(t)
=
1+\frac{3}{N}t
+
\frac{23-18/N}{4(N^2-1)}\,t^2
+O(t^3).
\label{eq:trajectory-quadratic}
\end{equation}
In particular,
\begin{equation}
F_N'(0)=\frac{3}{N}.
\label{eq:trajectory-slope}
\end{equation}

\begin{figure}[t]
    \centering
    \includegraphics[width=0.72\textwidth]
    {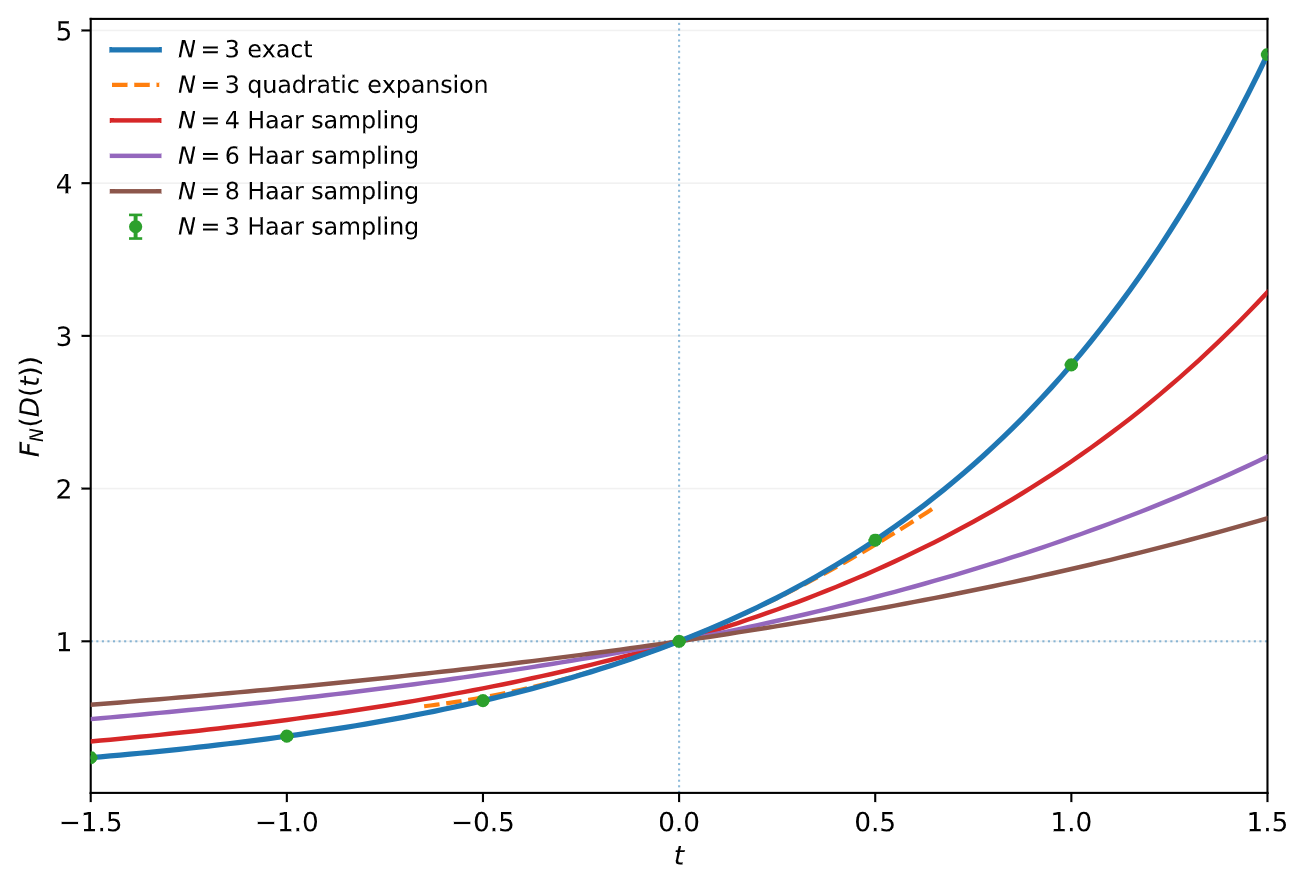}
    \caption{
    The reduced kernel along the genuinely rank-two source trajectory
    \eqref{eq:numerical-source}. For $t>0$ the nonzero $2\times2$ block
    satisfies $\det D=3t^2/4>0$ and therefore lies away from the rank-one
    HCIZ locus. The $t<0$ branch, where the nonzero $2\times2$ block of $D(t)$
    is negative definite, is shown as an analytic continuation off the physical
    conjugate locus. At $N=3$,
    direct Haar sampling agrees with the exact Bessel
    representation \eqref{eq:U3-Bessel}; the quadratic Weingarten expansion
    \eqref{eq:trajectory-quadratic} is shown near the origin. For
    $N=4,6,8$, the same rank-two kernel is evaluated directly by Haar
    sampling without requiring an explicit pushforward density.
    }
    \label{fig:haar-sampling}
\end{figure}

Figure~\ref{fig:haar-sampling} compares the exact $N=3$ result with direct
Haar sampling and extends the same rank-two computation to $N=4,6,8$.
The sampling curves were obtained from $5\times10^5$ Haar-distributed unitary
matrices for each value of $N$. For fixed $N$, the same Haar ensemble can be
used along the entire trajectory: defining
\begin{equation}
X(U)
=
|U_{11}|^2
+\frac12|U_{12}|^2
+\frac12|U_{21}|^2
+|U_{22}|^2,
\end{equation}
one has
\begin{equation}
F_N(t)=\left\langle e^{tX(U)}\right\rangle_H .
\label{eq:trajectory-expectation}
\end{equation}
At $N=3$, the Monte Carlo estimator agrees with the exact Bessel kernel within
statistical uncertainty, and its error decreases with the expected
$M^{-1/2}$ scaling. As expected, the quadratic Weingarten approximation
agrees near the origin and departs progressively from the exact kernel at
finite source, whereas Haar sampling continues to access the full Laplace
transform.

The finite-$N$ problem therefore separates naturally by source rank.
Rank-one sources are HCIZ-equivalent and admit exact analytic control for
arbitrary $N$, including the closed family
\eqref{eq:rank-one-hypergeometric}. For generic rank-two sources, the $N=3$
kernel remains exceptional in admitting an explicit pushforward reduction,
while at higher $N$ exact local data are available from Weingarten moments
and finite-source values from Haar sampling. This separation between exact
rank-one control and genuinely multi-matrix rank-two dynamics will be the
starting point for the quantum geometry developed in the following section.

\section{Finite-$N$ quantum geometry of the HCIZ embedding}
\label{sec:quantum-geometry}

\subsection{K\"ahler geometry from the coherent-state kernel}
\label{sec:kahler-from-kernel}

The reduced kernel contains a natural quantum-geometric structure in addition
to its interpretation as a generating function for Haar moments. On the
relative-coordinate slice obtained by fixing the reference eigenvalues
$a_N=b_N=0$, the elementary prefactor in
Eq.~\eqref{eq:kernel-factorization} vanishes,
\begin{equation}
R_N(C)=0,
\end{equation}
so that the coherent-state norm is simply $F_N(D)$. We may therefore use
\begin{equation}
K_N(D)=\log F_N(D)
\label{eq:kahler-potential-reduced}
\end{equation}
as a K\"ahler potential for the pullback of the Fubini--Study metric to this
reduced family of protected coherent states
\cite{provost1980riemannian}.

For the two-matrix problem, introduce relative holomorphic coordinates
\begin{equation}
\xi_{iX}=a_i-a_N,
\qquad
\xi_{iY}=b_i-b_N,
\qquad
i=1,\ldots,N-1,
\label{eq:relative-xi}
\end{equation}
so that on the physical conjugate locus
\begin{equation}
D_{ij}
=
\sum_{A=X,Y}
\xi_{iA}\bar\xi_{jA}.
\label{eq:D-xi}
\end{equation}
As in Section~\ref{sec:cumulants}, define
\begin{equation}
M_{ij}(D)
=
\frac{\partial K_N}{\partial D_{ij}}
=
\langle q_{ij}\rangle_D
\label{eq:M-def}
\end{equation}
and
\begin{equation}
\Gamma_{ij,kl}(D)
=
\frac{\partial^2 K_N}
{\partial D_{ij}\partial D_{kl}}
=
\operatorname{Cov}_D(q_{ij},q_{kl}),
\label{eq:Gamma-def}
\end{equation}
where the expectation values are taken in the exponentially tilted
unistochastic ensemble \eqref{eq:tilted-measure}. A direct application of the
chain rule then gives
\begin{equation}
g_{iA,\bar jB}
=
\delta_{AB}M_{ij}
+
\sum_{n,p=1}^{N-1}
\bar\xi_{nA}\xi_{pB}\,
\Gamma_{in,pj}.
\label{eq:general-pullback-metric}
\end{equation}
Thus the finite-$N$ coherent-state metric is determined by two pieces of
unistochastic data: the tilted mean of $q$ and its connected covariance.
Equation~\eqref{eq:general-pullback-metric} is the geometric counterpart of
the cumulant interpretation of $K_N$. With the barred variables kept
independent, as in Section~\ref{sec:unistochastic-reduction}, the logarithm of
the coherent-state norm is naturally of the overlap-generating-function form
$\log\langle\Psi(s')|\Psi(s)\rangle$ used in
Ref.~\cite{HetenyiLevay2023}. In that
language, Eq.~\eqref{eq:general-pullback-metric} is its second-order cumulant
with respect to the coherent-state coordinates $\xi,\bar\xi$, and the
connection and curvature used in Section~\ref{sec:gauss-codazzi} are built
from its third- and fourth-order cumulants. Because $D$ depends quadratically
on $\xi$, each of these is assembled through the chain rule from source
cumulants of the tilted ensemble ($M$, $\Gamma$, and their higher-order
analogues) rather than coinciding with a single source cumulant.

At the rank-zero point $\xi=0$, only the first term survives. Haar symmetry
gives
\begin{equation}
M_{ij}(0)=\frac{1}{N},
\end{equation}
so the metric on each scalar block is proportional to the matrix with all
entries equal to $1/N$. The pullback metric is therefore degenerate at the
apex of the reduced source cone. This degeneracy is a property of the
coherent-state map rather than of the coordinates: at zero source the tilted
mean is the van der Waerden matrix, which has rank one. Equivalently, for the
coherent states
\begin{equation}
|\Lambda_X,\Lambda_Y\rangle
=
\frac{1}{\operatorname{Vol}[U(N)]}
\int dU\,
\exp\left(
\operatorname{Tr}\left[
UXU^\dagger\Lambda_X
+
UYU^\dagger\Lambda_Y
\right]\right)|0\rangle
\end{equation}
of Ref.~\cite{holguin2024multi}, all first-order variations at the apex
coincide,
\begin{equation}
\partial_{a_i}|\Lambda_X,\Lambda_Y\rangle\big|_{\Lambda=0}
=
\frac1N\operatorname{Tr}X\,|0\rangle,
\qquad
\partial_{b_i}|\Lambda_X,\Lambda_Y\rangle\big|_{\Lambda=0}
=
\frac1N\operatorname{Tr}Y\,|0\rangle,
\label{eq:apex-variations}
\end{equation}
for every $i$, since $\int d\mu_H(U)\,U_{ik}\bar U_{il}=\delta_{kl}/N$. The
map from relative coordinates to projective Hilbert space therefore fails to
be an immersion at the apex, even though the projective Hilbert space itself
is smooth.

\subsection{The first nontrivial case: $N=3$}
\label{sec:N3-quantum-geometry}

For $N=3$, ordering the relative coordinates as
\begin{equation}
(\alpha_1,\alpha_2,\beta_1,\beta_2),
\end{equation}
Eq.~\eqref{eq:general-pullback-metric} gives at the apex
\begin{equation}
g(0)
=
\frac13
\begin{pmatrix}
1&1&0&0\\
1&1&0&0\\
0&0&1&1\\
0&0&1&1
\end{pmatrix},
\label{eq:N3-vacuum-metric}
\end{equation}
with eigenvalues
\begin{equation}
\operatorname{spec}g(0)
=
\left\{
\frac23,\frac23,0,0
\right\}.
\label{eq:N3-vacuum-spectrum}
\end{equation}
The two null directions are relative combinations that are invisible to the
kernel at linear order around the rank-zero configuration and acquire
nonzero norm only away from the apex.

The exact $N=3$ kernel makes this opening of the geometry explicit. Along the
rank-one ray
\begin{equation}
\alpha=(a,0),
\qquad
\beta=(0,0),
\qquad
r=|a|^2,
\end{equation}
Eq.~\eqref{eq:U3-rank-one-ray} gives
\begin{equation}
F_3(rE_{11})
=
\frac{2(e^r-1-r)}{r^2}.
\end{equation}
Writing
\begin{equation}
k_3(r)=\log F_3(rE_{11}),
\qquad
\mu_3(r)=k_3'(r),
\qquad
s_3(r)=k_3''(r),
\end{equation}
the radial metric is
\begin{equation}
g_{a\bar a}^{(3)}
=
\mu_3+r s_3
=
\frac{
e^{2r}-2e^r+1-r^2e^r
}{
(e^r-1-r)^2
}.
\label{eq:N3-radial-metric}
\end{equation}
Near the apex,
\begin{equation}
g_{a\bar a}^{(3)}
=
\frac13+\frac{r}{9}
+\frac{r^2}{90}
-\frac{r^3}{810}
+O(r^4).
\label{eq:N3-radial-series}
\end{equation}
The radial metric is therefore positive for all $r\geq0$, with
$g^{(3)}_{a\bar a}(0)=1/3$, even though the full relative-coordinate pullback
is degenerate at $r=0$. The exact Bessel kernel
may likewise be differentiated at generic rank-two points, but the
rank-one family admits a stronger result: its normal geometry can be
determined analytically for arbitrary finite $N$.

\subsection{Normal geometry of the rank-one locus at arbitrary $N$}
\label{sec:normal-geometry-general-N}

We now probe how the HCIZ rank-one stratum sits inside the two-matrix
coherent-state geometry. We restrict to the
three-complex-dimensional family
\begin{equation}
\alpha=(a,0,\ldots,0),
\qquad
\beta=(u,v,0,\ldots,0),
\qquad
r=|a|^2,
\label{eq:controlled-family}
\end{equation}
for $N\geq3$. On the physical conjugate locus,
\begin{equation}
D
=
\begin{pmatrix}
r+|u|^2 & u\bar v\\
v\bar u & |v|^2
\end{pmatrix}
\oplus\mathbf 0_{N-3},
\label{eq:controlled-D}
\end{equation}
and hence
\begin{equation}
\det D_{\{1,2\}}
=
r|v|^2.
\label{eq:controlled-det}
\end{equation}
For $r>0$, the surface
\begin{equation}
\mathcal H:\qquad v=0
\label{eq:H-surface}
\end{equation}
lies in the rank-one HCIZ locus, while $v$ is a representative genuinely
rank-two transverse deformation.

\paragraph{Choice of slice.}
The background $u=v=0$ corresponds to coincident eigenvalues,
$a_2=\cdots=a_N$ and $b_1=\cdots=b_N$. For $N\geq4$ this makes the pullback
metric \eqref{eq:general-pullback-metric} on the full space of relative
coordinates $(\xi_{iX},\xi_{iY})\in\mathbb C^{2(N-1)}$ degenerate along the
entire ray. Because the tilted ensemble is invariant under permutations of
rows and columns $2,\ldots,N$, the one-point functions $M_{ij}$ and the
covariances $\Gamma_{i1,1j}$ do not depend on $i,j\geq2$. The variations
$\delta\xi_{\cdot A}\propto e_j-e_k$ with $2\leq j<k\leq N-1$ are therefore
null for both $A=X,Y$. Some of these directions, for example
$\delta\beta\propto e_2-e_3$, are genuinely rank-two deformations. This is
the familiar failure of the coherent-state map to be an immersion at
coincident eigenvalues: first-order variations that are traceless within a
degenerate eigenvalue block average to zero over its stabilizer. The family
\eqref{eq:controlled-family} avoids these directions, and all statements
below about the normal geometry refer to $\mathcal H$ inside this family.
For $N=3$ no such null directions exist, and the $(a,u)$ component of the
second fundamental form computed below coincides with the corresponding
component for the rank-one hypersurface $\det D=0$ in the full
four-complex-dimensional $N=3$ geometry
(Appendix~\ref{app:full-space-N3}).

Along the background $u=v=0$ the source is $D=rE_{11}$, whose kernel is the
exact all-$N$ expression \eqref{eq:rank-one-hypergeometric}. Define
\begin{equation}
k_N(r)
=
\log F_N(rE_{11}),
\qquad
\mu_N(r)=k_N'(r),
\qquad
s_N(r)=k_N''(r).
\label{eq:mu-s-def}
\end{equation}
The first two derivatives have the probabilistic interpretation
\begin{equation}
\mu_N(r)=\langle q_{11}\rangle_r,
\qquad
s_N(r)=\operatorname{Var}_r(q_{11})>0,
\label{eq:mu-s-prob}
\end{equation}
and
\begin{equation}
\mu_N(r)
=
\frac{1}{N}
\frac{
{}_1F_1(2;N+1;r)
}{
{}_1F_1(1;N;r)
}.
\label{eq:mu-hypergeom}
\end{equation}

Because the tilted weight depends only on $q_{11}$, permutations of rows and
columns $2,\ldots,N$ remain symmetries of the ensemble. Double stochasticity
then fixes the relevant one-point functions to
\begin{equation}
M_{11}=\mu_N,
\qquad
M_{1j}=M_{j1}
=
\frac{1-\mu_N}{N-1},
\qquad
j>1,
\label{eq:M-ray-first}
\end{equation}
and
\begin{equation}
M_{ij}
=
\frac{N-2+\mu_N}{(N-1)^2},
\qquad
i,j>1.
\label{eq:M-ray-rest}
\end{equation}
At $u=v=0$, Eq.~\eqref{eq:general-pullback-metric} therefore gives, in the
coordinates $(a,u,v)$,
\begin{equation}
g
=
\begin{pmatrix}
A_N & 0 & 0\\[1mm]
0 & \mu_N & m_N\\[1mm]
0 & m_N & d_N
\end{pmatrix},
\label{eq:controlled-metric}
\end{equation}
where
\begin{equation}
A_N=\mu_N+r s_N,
\qquad
m_N=\frac{1-\mu_N}{N-1},
\qquad
d_N=\frac{N-2+\mu_N}{(N-1)^2}.
\label{eq:Amd-def}
\end{equation}

The algebraic transverse vector $\partial_v$ is not orthogonal to the
rank-one tangent direction $\partial_u$. The metric-orthogonal lift of the
holomorphic quotient frame $e=[\partial_v]$ is
\begin{equation}
n
=
\partial_v
-
\frac{1-\mu_N}{(N-1)\mu_N}\,
\partial_u .
\label{eq:normal-vector}
\end{equation}
Its squared norm, equivalently the Schur complement of the tangent block, is
\begin{equation}
h_N(r)
=
\|n\|^2
=
\frac{
N\mu_N(r)-1
}{
(N-1)^2\mu_N(r)
}.
\label{eq:normal-metric}
\end{equation}
Since $\mu_N(0)=1/N$ and $\mu_N'(r)=s_N(r)>0$, one has
$h_N(r)>0$ for every physical $r>0$. Within the controlled family, the HCIZ
surface is therefore not separated from the rank-two region by a singular
wall in this finite-$N$ fidelity geometry: the representative second-scalar
deformation $\partial_v$ has finite projective length at every nonzero
rank-one background. For $N\geq4$ this does not extend to all rank-two
directions at this background, as explained below
Eq.~\eqref{eq:H-surface}.

Near the apex,
\begin{equation}
\mu_N(r)
=
\frac1N
+
\frac{N-1}{N^2(N+1)}\,r
+
O(r^2),
\label{eq:mu-small-r}
\end{equation}
and hence
\begin{equation}
h_N(r)
=
\frac{r}{N^2-1}
+
O(r^2).
\label{eq:h-small-r}
\end{equation}
Thus the normal direction collapses only at the rank-zero configuration.
For fixed $N$ and large source,
\begin{equation}
\mu_N(r)
=
1-\frac{N-1}{r}
+O\!\left(r^{N-2}e^{-r}\right),
\end{equation}
so that
\begin{equation}
h_N(r)
=
\frac{1}{N-1}
\left(
1-\frac{1}{r-N+1}
\right)
+O\!\left(r^{N-2}e^{-r}\right)
\longrightarrow
\frac{1}{N-1}.
\label{eq:h-large-r-fixed-N}
\end{equation}

\subsection{Second fundamental form and Gauss--Codazzi geometry}
\label{sec:gauss-codazzi}

The normal metric probes the transverse geometry of the smooth rank-one locus, but does not determine whether the locus is geodesically closed inside the larger two-matrix geometry. Let $e=[\partial_v]$ denote the holomorphic quotient frame of the
normal line to $\mathcal H$ in the controlled family
\eqref{eq:controlled-family}. A direct evaluation of the ambient K\"ahler
connection at $u=v=0$ gives
\begin{equation}
B_{aa}=B_{uu}=0,
\end{equation}
while the mixed component is
\begin{equation}
B_{au}=B_{ua}
=
-
\frac{
(N-1)\bar a\,s_N(r)
}{
N\mu_N(r)-1
}\,e .
\label{eq:second-fundamental-form}
\end{equation}
For every finite $N$ and $r>0$, $s_N>0$, so this component is nonzero.
The HCIZ surface is therefore not totally geodesic in the controlled family:
the controlled slice exhibits an explicit tangent pair whose ambient
covariant derivative has a genuine rank-two normal component. The invariant
norm below is the norm on the two-complex-dimensional controlled surface
$\mathcal H$, evaluated at $u=v=0$.\footnote{All extrinsic quantities in this
subsection are evaluated at $u=v=0$. The controlled family is not invariant
under the $U(2)$ rotation mixing $a$ and $u$, since that rotation does not
preserve the $v$ direction, so their values at $u\neq0$ are not determined by
$|a|^2+|u|^2$ alone.}

Using the normal metric \eqref{eq:normal-metric},
\begin{equation}
\|B_{au}\|^2
=
\frac{
r\,s_N(r)^2
}{
\mu_N(r)\left(N\mu_N(r)-1\right)
},
\label{eq:Bau-norm}
\end{equation}
and the full squared norm of the second fundamental form of the
two-complex-dimensional HCIZ surface $\mathcal H$ at $u=v=0$ is
\begin{equation}
\|B\|^2
=
\frac{
2r\,s_N(r)^2
}{
\left(\mu_N+r s_N\right)
\mu_N^2
\left(N\mu_N-1\right)
}.
\label{eq:B-full-norm}
\end{equation}
Near the rank-zero apex,
\begin{equation}
\|B\|^2
\longrightarrow
\frac{2(N-1)}{N+1},
\qquad
r\to0^+,
\label{eq:B-small-r}
\end{equation}
whereas for fixed $N$ at large source,
\begin{equation}
\|B\|^2
\sim
\frac{2(N-1)}{r^3}.
\label{eq:B-large-r-fixed-N}
\end{equation}
The coordinate coefficient in
Eq.~\eqref{eq:second-fundamental-form} is singular as $r\to0$, because the
normal frame itself collapses there, but the invariant norm has a finite
one-sided limit. Exactly at $r=0$ the rank-one stratum meets the rank-zero
apex and the smooth-submanifold description used here ceases to apply.

The curvature comparison makes the role of this extrinsic bending explicit.
We use the convention
\begin{equation}
R_{i\bar j k\bar l}
=
-\partial_i\partial_{\bar j}g_{k\bar l}
+
g^{p\bar q}
(\partial_i g_{k\bar q})
(\partial_{\bar j}g_{p\bar l}).
\label{eq:curvature-convention}
\end{equation}
On $\mathcal H$, the induced K\"ahler potential is
\begin{equation}
K_{\mathcal H}
=
k_N\!\left(|a|^2+|u|^2\right).
\label{eq:H-kahler-potential}
\end{equation}
Writing
\begin{equation}
t_N(r)=s_N'(r)=k_N'''(r),
\end{equation}
the intrinsic mixed curvature at $u=0$ is
\begin{equation}
R^{\mathcal H}_{a\bar a u\bar u}
=
-s_N-r t_N
+
\frac{r s_N^2}{\mu_N}.
\label{eq:intrinsic-mixed-curvature}
\end{equation}
The corresponding ambient curvature component is
\begin{equation}
R^{\mathcal M}_{a\bar a u\bar u}
=
-s_N-r t_N
+
\frac{
N r s_N^2
}{
N\mu_N-1
}.
\label{eq:ambient-mixed-curvature}
\end{equation}
Their difference is precisely
\begin{equation}
R^{\mathcal M}_{a\bar a u\bar u}
-
R^{\mathcal H}_{a\bar a u\bar u}
=
\frac{
r s_N^2
}{
\mu_N(N\mu_N-1)
}
=
\|B_{au}\|^2,
\label{eq:gauss-mixed}
\end{equation}
which is the mixed component of the Gauss equation. Since
$B_{aa}=B_{uu}=0$, the purely radial and purely $u$-direction tangent
curvatures coincide intrinsically and ambiently; the difference appears
precisely in the mixed plane in which the embedding bends toward the
rank-two direction.

Equivalently, if
\begin{equation}
S_{\mathcal M}^{\parallel}
=
g^{i\bar j}g^{k\bar l}
R^{\mathcal M}_{i\bar j k\bar l}
\end{equation}
denotes the ambient curvature traced only over tangent directions to
$\mathcal H$, then at $u=v=0$
\begin{equation}
S_{\mathcal M}^{\parallel}
-
S_{\mathcal H}
=
\|B\|^2.
\label{eq:scalar-gauss}
\end{equation}
Part of the intrinsic curvature of the HCIZ family is therefore extrinsic:
it is generated by the way the rank-one state family bends inside the
larger two-matrix coherent-state geometry.

The normal line bundle also has a closed finite-$N$ curvature. With the
convention
\begin{equation}
\Theta^\perp
=
-\partial\bar\partial\log h_N,
\end{equation}
its radial component is
\begin{equation}
\Theta^\perp_{a\bar a}
=
-
\frac{
s_N+r t_N
}{
\mu_N(N\mu_N-1)
}
+
\frac{
r s_N^2(2N\mu_N-1)
}{
\mu_N^2(N\mu_N-1)^2
}.
\label{eq:normal-bundle-curvature}
\end{equation}
The corresponding Codazzi relation may be written as
\begin{equation}
\left(
R^{\mathcal M}_{a\bar a}\partial_u
\right)^\perp
=
-
\left(
\nabla^\perp_{\bar a}B
\right)_{au}
=
\mathcal C_N(r)\,e,
\label{eq:codazzi-relation}
\end{equation}
with
\begin{equation}
\mathcal C_N(r)
=
(N-1)
\frac{
(s_N+r t_N)(N\mu_N-1)-N r s_N^2
}{
(N\mu_N-1)^2
}.
\label{eq:codazzi-coefficient}
\end{equation}

As a representative finite-source example, at $N=3$ and $r=1$ one finds
\begin{equation}
\mu_3=0.39221119,
\qquad
s_3=0.06174800,
\qquad
h_3=0.11258831,
\end{equation}
and
\begin{equation}
\|B\|^2=0.61822383.
\end{equation}
The mixed intrinsic and ambient curvature components are
\begin{equation}
R^{\mathcal H}_{a\bar a u\bar u}
=
-0.05672773,
\qquad
R^{\mathcal M}_{a\bar a u\bar u}
=
-0.00169101.
\end{equation}
After normalizing by
$g_{a\bar a}g_{u\bar u}$, the corresponding bisectional curvatures are
approximately
\begin{equation}
-0.31861
\qquad\text{and}\qquad
-0.00950,
\end{equation}
respectively. At this point most of the negative mixed curvature seen
intrinsically on the HCIZ surface is therefore accounted for by its
extrinsic bending rather than by the curvature of the corresponding tangent
plane in the ambient two-matrix geometry.

The fixed-$N$ results above already show that the HCIZ sector is a smooth but
non-totally-geodesic lower-rank family inside the controlled two-matrix
family away from the rank-zero apex, and at $N=3$ also inside the full
two-matrix coherent-state geometry. The remaining question is how this
embedding behaves when the coherent-state source and $N$ are scaled
simultaneously. We turn to that controlled large-$N$ limit in the next
section.

\section{Controlled large-$N$ geometry}
\label{sec:large-N-geometry}

The exact rank-one family obtained in
Eq.~\eqref{eq:rank-one-hypergeometric} admits a controlled large-$N$
limit in which the source is scaled together with the matrix size. This
limit is distinct from the fixed-$N$, large-$r$ regime discussed in
Section~\ref{sec:normal-geometry-general-N}. Since the source rank remains finite, the natural free-energy
normalization is $N^{-1}$ rather than $N^{-2}$, as in the general theory
of finite-rank spherical integrals and, in particular, rank-one HCIZ
asymptotics
\cite{guionnet2005fourier,mergny2022rank}. Here the beta-integral representation makes the
relevant saddle structure completely explicit and allows us to propagate it
directly into the normal and extrinsic geometry of the HCIZ embedding.

\subsection{Boundary and interior saddles on the rank-one ray}
\label{sec:large-N-saddle}

We scale
\begin{equation}
r=N\rho,
\qquad
\rho>0,
\label{eq:large-N-scaling}
\end{equation}
while keeping the reduced source rank equal to one. Equation
\eqref{eq:rank-one-hypergeometric} becomes
\begin{equation}
F_N(N\rho E_{11})
=
(N-1)
\int_0^1 dq\,
\exp\left[
N\rho q+(N-2)\log(1-q)
\right].
\label{eq:large-N-beta-integral}
\end{equation}
At leading exponential order the relevant function is
\begin{equation}
\phi_\rho(q)
=
\rho q+\log(1-q),
\qquad
0\leq q<1.
\label{eq:large-N-rate-function}
\end{equation}
Its derivative is
\begin{equation}
\phi_\rho'(q)
=
\rho-\frac{1}{1-q}.
\end{equation}
For $0<\rho<1$, $\phi_\rho$ decreases from the endpoint $q=0$, so the
integral is controlled by a boundary saddle. For $\rho>1$, an interior
stationary point appears at
\begin{equation}
q_\star
=
1-\frac{1}{\rho}.
\label{eq:interior-saddle}
\end{equation}
The two regimes meet at $\rho=1$, where the boundary saddle becomes
quadratically degenerate.

Define the finite-rank free energy
\begin{equation}
\Phi_N(\rho)
=
\frac{1}{N}
\log F_N(N\rho E_{11}).
\label{eq:PhiN-def}
\end{equation}
Laplace's method gives
\begin{equation}
\Phi(\rho)
=
\lim_{N\to\infty}\Phi_N(\rho)
=
\begin{cases}
0,
&
0<\rho\leq1,
\\[2mm]
\rho-1-\log\rho,
&
\rho>1.
\end{cases}
\label{eq:large-N-free-energy}
\end{equation}
The limiting free energy is continuously differentiable at $\rho=1$ but has
a discontinuity in its second derivative. We will therefore refer to
$\rho=1$ as a boundary-to-interior saddle transition rather than infer a
separate dynamical phase transition of the underlying gauge theory.

The subleading asymptotics also make the change in the tilted ensemble
transparent. For fixed $0<\rho<1$, set $q=x/N$ in
Eq.~\eqref{eq:large-N-beta-integral}. Then
\begin{equation}
F_N(N\rho E_{11})
=
\frac{1}{1-\rho}
\left[
1+O(N^{-1})
\right],
\qquad
0<\rho<1,
\label{eq:F-below-threshold}
\end{equation}
and the tilted variable remains of order $N^{-1}$,
\begin{equation}
q_{11}=O(N^{-1}).
\end{equation}
For $\rho>1$, the interior saddle
Eq.~\eqref{eq:interior-saddle} instead gives
\begin{equation}
F_N(N\rho E_{11})
=
\rho\sqrt{2\pi N}\,
\exp\left[
N\bigl(\rho-1-\log\rho\bigr)
\right]
\left[
1+O(N^{-1})
\right],
\label{eq:F-above-threshold}
\end{equation}
so the tilted ensemble develops an order-one value of $q_{11}$ centered at
$q_\star$.

Exactly at the transition, the appropriate scale is
\begin{equation}
q=N^{-1/2}x.
\end{equation}
More generally, in the critical window
\begin{equation}
\rho
=
1+\frac{\lambda}{\sqrt N},
\qquad
\lambda=O(1),
\label{eq:critical-window}
\end{equation}
one obtains the uniform leading form
\begin{equation}
F_N(N\rho E_{11})
=
\sqrt N\,
\Psi(\lambda)
\left[
1+O(N^{-1/2})
\right],
\label{eq:critical-F}
\end{equation}
where
\begin{equation}
\Psi(\lambda)
=
\int_0^\infty
\exp\left(
\lambda x-\frac{x^2}{2}
\right)
dx.
\label{eq:Psi-critical}
\end{equation}
In particular,
\begin{equation}
F_N(NE_{11})
\sim
\sqrt{\frac{\pi N}{2}},
\label{eq:F-critical}
\end{equation}
and $q_{11}=O(N^{-1/2})$ at $\rho=1$. The three characteristic scales of
the tilted variable are therefore
\begin{equation}
q_{11}
\sim
\begin{cases}
N^{-1},
&
0<\rho<1,
\\
N^{-1/2},
&
\rho=1,
\\
O(1),
&
\rho>1.
\end{cases}
\label{eq:q-scaling-regimes}
\end{equation}

\subsection{Large-$N$ normal metric}
\label{sec:large-N-normal-metric}

The saddle asymptotics determine the geometry through the derivatives
introduced in Eq.~\eqref{eq:mu-s-def},
\begin{equation}
\mu_N(r)=\partial_r\log F_N(rE_{11}),
\qquad
s_N(r)=\partial_r^2\log F_N(rE_{11}).
\end{equation}
For fixed $0<\rho<1$, Eq.~\eqref{eq:F-below-threshold} gives
\begin{equation}
\mu_N(N\rho)
=
\frac{1}{N(1-\rho)}
+
O(N^{-2}),
\label{eq:mu-below}
\end{equation}
and
\begin{equation}
s_N(N\rho)
=
\frac{1}{N^2(1-\rho)^2}
+
O(N^{-3}).
\label{eq:s-below}
\end{equation}
For fixed $\rho>1$, the interior saddle gives
\begin{equation}
\mu_N(N\rho)
=
1-\frac{1}{\rho}
+
O(N^{-1}),
\label{eq:mu-above}
\end{equation}
and
\begin{equation}
s_N(N\rho)
=
\frac{1}{N\rho^2}
+
O(N^{-2}).
\label{eq:s-above}
\end{equation}
Thus $\mu_N$, which is the tilted expectation value
$\langle q_{11}\rangle$, directly records the change from a microscopic to
a macroscopic matrix element.

Substituting these expressions into the exact finite-$N$ normal metric
\eqref{eq:normal-metric} gives a finite result after rescaling by $N$:
\begin{equation}
\lim_{N\to\infty}
N\,h_N(N\rho)
=
\begin{cases}
\rho,
&
0<\rho<1,
\\[1mm]
1,
&
\rho\geq1.
\end{cases}
\label{eq:large-N-normal-metric-limit}
\end{equation}
Equivalently,
\begin{equation}
N\,h_N(N\rho)
\longrightarrow
\min(\rho,1).
\label{eq:large-N-normal-metric-min}
\end{equation}
The physical normal metric itself is therefore of order $N^{-1}$ throughout
the double-scaling regime, but its rescaled coefficient retains a sharp
memory of the saddle transition.

This result should be contrasted with the fixed-$N$, large-source limit
\eqref{eq:h-large-r-fixed-N}, where
$h_N\to1/(N-1)$. The two limits agree on the $\rho>1$ side after taking
$N\to\infty$, but the $\rho<1$ regime is nonuniform: even though
$r=N\rho\to\infty$, the entropy of the Haar measure keeps the dominant
$q_{11}$ near the boundary and reduces the rescaled normal metric by the
factor $\rho$.

The critical window \eqref{eq:critical-window} smooths the cusp at finite
$N$. Defining
\begin{equation}
m(\lambda)
=
\frac{d}{d\lambda}\log\Psi(\lambda),
\qquad
v(\lambda)
=
\frac{d^2}{d\lambda^2}\log\Psi(\lambda),
\label{eq:critical-m-v}
\end{equation}
one finds
\begin{equation}
\mu_N
=
\frac{m(\lambda)}{\sqrt N}
+
O(N^{-1}),
\qquad
s_N
=
\frac{v(\lambda)}{N}
+
O(N^{-3/2}),
\label{eq:critical-mu-s}
\end{equation}
Using the exact identity
\begin{equation}
N h_N
=
\left(\frac{N}{N-1}\right)^2
\left(
1-\frac{1}{N\mu_N}
\right),
\label{eq:Nh-identity}
\end{equation}
which follows directly from Eq.~\eqref{eq:normal-metric}, one finds
\begin{equation}
N h_N
=
1-\frac{1}{\sqrt N\,m(\lambda)}
+O(N^{-1})
\label{eq:critical-h}
\end{equation}
throughout the $N^{-1/2}$ critical window. Since
$m(\lambda)=-1/\lambda+O(\lambda^{-3})$ as $\lambda\to-\infty$, while
$m(\lambda)-\lambda$ is exponentially small as $\lambda\to+\infty$,
Eq.~\eqref{eq:critical-h} interpolates between $Nh_N\simeq\rho$ and
$Nh_N\simeq1$. This is how the cusp in
Eq.~\eqref{eq:large-N-normal-metric-min} is smoothed at finite $N$.

\subsection{Extrinsic curvature across the saddle transition}
\label{sec:large-N-extrinsic}

The large-$N$ behavior of the second fundamental form is more strongly
regime dependent. Substituting Eqs.~\eqref{eq:mu-below} and
\eqref{eq:s-below} into the exact expression
\eqref{eq:B-full-norm} yields
\begin{equation}
\lim_{N\to\infty}
\|B\|^2_{r=N\rho}
=
2(1-\rho),
\qquad
0<\rho<1.
\label{eq:B-large-N-below}
\end{equation}
On the interior-saddle side,
\begin{equation}
\|B\|^2_{r=N\rho}
=
\frac{2}{
N^2(\rho-1)^3
}
+
O(N^{-3}),
\qquad
\rho>1,
\label{eq:B-large-N-above}
\end{equation}
and therefore
\begin{equation}
\lim_{N\to\infty}
\|B\|^2_{r=N\rho}
=
2(1-\rho)_+,
\label{eq:B-large-N-limit}
\end{equation}
where $(x)_+=\max(x,0)$.

Thus the controlled rank-one slice remains extrinsically curved at leading
order for $0<\rho<1$, while it becomes asymptotically totally geodesic in
the representative rank-two normal direction for $\rho>1$. At the
transition itself, Eq.~\eqref{eq:critical-mu-s} gives
\begin{equation}
\|B\|^2
=
\frac{
2v(\lambda)
}{
m(\lambda)^3
}
N^{-1/2}
+
O(N^{-1})
\label{eq:B-critical-window}
\end{equation}
within the critical window, so the two regimes join continuously. Indeed,
$2v/m^3\to2|\lambda|$ as $\lambda\to-\infty$ and $2v/m^3\to2/\lambda^3$ as
$\lambda\to+\infty$, which reproduce Eqs.~\eqref{eq:B-large-N-below} and
\eqref{eq:B-large-N-above}, respectively.

The same structure appears directly in the Gauss equation. Let
\begin{equation}
\mathcal K_{\mathcal H}
=
\frac{
R^{\mathcal H}_{a\bar a u\bar u}
}{
g_{a\bar a}g_{u\bar u}
},
\qquad
\mathcal K_{\mathcal M}
=
\frac{
R^{\mathcal M}_{a\bar a u\bar u}
}{
g_{a\bar a}g_{u\bar u}
},
\label{eq:normalized-bisectional}
\end{equation}
denote the intrinsic and ambient mixed bisectional curvatures on the
controlled HCIZ surface. For fixed $0<\rho<1$, the asymptotic forms of
$\mu_N$, $s_N$, and $t_N=s_N'$ in
Eqs.~\eqref{eq:intrinsic-mixed-curvature} and
\eqref{eq:ambient-mixed-curvature} give
\begin{equation}
\mathcal K_{\mathcal H}
\longrightarrow
-1,
\qquad
\mathcal K_{\mathcal M}
\longrightarrow
-\rho,
\qquad
0<\rho<1.
\label{eq:bisectional-below}
\end{equation}
Consequently,
\begin{equation}
\mathcal K_{\mathcal M}
-
\mathcal K_{\mathcal H}
\longrightarrow
1-\rho,
\label{eq:bisectional-gap}
\end{equation}
which is precisely the normalized contribution of the second fundamental
form required by Eq.~\eqref{eq:gauss-mixed}.

For fixed $\rho>1$, both normalized curvatures vanish:
\begin{equation}
\begin{split}
\mathcal K_{\mathcal H}
&=
\frac{1}{N(\rho-1)^2}
+O(N^{-2}),
\\
\mathcal K_{\mathcal M}
-
\mathcal K_{\mathcal H}
&=
\frac12\|B\|^2
=
\frac{1}{N^2(\rho-1)^3}
+O(N^{-3}).
\end{split}
\label{eq:bisectional-above}
\end{equation}
The intrinsic and ambient tangent geometries therefore become asymptotically
indistinguishable on the interior-saddle side, with both mixed curvatures
positive at leading order.

Taken together, Eqs.~\eqref{eq:large-N-normal-metric-min} and
\eqref{eq:B-large-N-limit} give a compact description of the controlled
large-$N$ embedding:
\begin{equation}
N h_N(N\rho)
\longrightarrow
\min(\rho,1),
\qquad
\|B\|^2
\longrightarrow
2(1-\rho)_+.
\label{eq:large-N-geometric-summary}
\end{equation}
The first quantity measures the rescaled projective cost of leaving the
HCIZ sector in the representative second-scalar direction, while the second
measures the extrinsic bending of that sector inside the controlled
two-matrix family. Below the saddle threshold the normal direction remains
soft, but the rank-one family retains finite extrinsic curvature. Above the
threshold the normal metric saturates after rescaling and the extrinsic
bending disappears at leading order.

\subsection{Interpretation and relation to finite-rank spherical integrals}
\label{sec:large-N-interpretation}

The free-energy behavior in Eq.~\eqref{eq:large-N-free-energy} is a
finite-rank spherical-integral phenomenon: an external source of order $N$
competes with the Haar entropy of a single matrix element. The resulting
boundary-to-interior saddle transition is therefore not by itself a new
large-$N$ phase of the gauge theory. What is specific to the present
coherent-state problem is that the same transition controls the embedding
geometry between source-rank strata.

In particular, the rank-one HCIZ family does not become uniformly
geodesically autonomous merely because $N$ is large. In the controlled
slice, its extrinsic curvature survives at order one for
$0<\rho<1$ and vanishes only as the interior saddle is reached. For
$\rho>1$, the rank-one family becomes asymptotically geodesic in the
representative rank-two normal direction, while the intrinsic and ambient
mixed curvatures simultaneously collapse. This provides a geometric sense
in which the effectively one-scalar collective-coordinate sector becomes
increasingly autonomous at sufficiently strong scaled source, without
implying a dynamical decoupling or stability statement for individual
giant-graviton states.

The rank-one saddle asymptotics themselves are consistent with the known
large-$N$ theory of finite-rank spherical integrals
\cite{guionnet2005fourier}; the additional result here is the induced
finite-$N$ and large-$N$ normal geometry of the rank-stratified BPS
coherent-state family. Together with the exact $N=3$ rank-two kernel and the
arbitrary-$N$ finite-rank formulas of the preceding sections, this completes
the analytic picture needed for the discussion and conclusions.

\section{Discussion}
\label{sec:conclusion}

The results of this paper reorganize the finite-$N$ multi-matrix
coherent-state kernel around two related structures: the Haar pushforward to
unistochastic variables and the rank geometry of the reduced source. After
simultaneous diagonalization of the commuting coherent-state data, the
unitary matrix enters the kernel only through
\begin{equation}
q_{ij}=|U_{ij}|^2,
\end{equation}
so the Haar integral descends exactly to a Laplace transform over the
unistochastic set. At the same time, double stochasticity removes the
row- and column-shift redundancy of the source and isolates the
$(N-1)^2$ invariant combinations
\begin{equation}
D_{ij}
=
C_{ij}-C_{iN}-C_{Nj}+C_{NN}.
\end{equation}
The kernel therefore factorizes as
\begin{equation}
\mathcal I_N(C)
=
e^{R_N(C)}F_N(D),
\end{equation}
with the nontrivial dynamics contained in
\begin{equation}
F_N(D)
=
\int_{\mathcal U_N}
\exp\!\left(
\sum_{i,j=1}^{N-1}D_{ij}q_{ij}
\right)d\nu_N(q).
\end{equation}

The additional observation developed here is that the physical
coherent-state locus is naturally stratified by the rank of $D$. For $m$
commuting scalar matrices,
\begin{equation}
D_{ij}
=
\sum_{A=1}^{m}
\xi_{iA}\bar\xi_{jA},
\end{equation}
on the physical conjugate locus, so that $D$ is positive semidefinite and
\begin{equation}
\operatorname{rank}D\leq m.
\end{equation}
The rank therefore counts the number of linearly independent relative scalar
directions probed by the source, rather than simply the number of scalar
matrices present. In particular, the ordinary HCIZ problem is precisely the
rank-at-most-one stratum modulo the row- and column-shift redundancy:
a reduced source admits an ordinary separable HCIZ representative if and
only if
\begin{equation}
\operatorname{rank}D\leq1.
\end{equation}
This characterization separates effectively one-scalar configurations from
genuinely multi-matrix source geometry in a basis-independent way.

The distinction first becomes nontrivial at $N=3$. For $N=2$ the reduced
source is a single scalar and is therefore automatically rank at most one.
At $N=3$ it is a $2\times2$ matrix and may have rank two. On the physical
two-matrix locus,
\begin{equation}
D
=
\alpha\alpha^\dagger+\beta\beta^\dagger,
\end{equation}
and
\begin{equation}
\det D
=
\left|
\alpha_1\beta_2-\alpha_2\beta_1
\right|^2.
\end{equation}
Thus the physical source space fills the cone $\operatorname{Herm}_2^+$:
the apex has rank zero, the nonzero boundary has rank one and is
HCIZ-equivalent, and the interior is genuinely rank two. The determinant is
the Gram determinant of the two relative scalar configurations and measures
their departure from an effectively one-dimensional scalar configuration.

The exact $N=3$ Bessel representation derived from the Haar pushforward is
therefore more than a compact rewriting of the direct calculation. It is the
first exact kernel that probes both the HCIZ rank-one stratum and a genuinely
rank-two source region within the same finite-$N$ expression. The
unistochastic discriminant $Q$ and the source determinant $\det D$ describe
different geometries: $Q$ determines the integration domain, whereas
$\operatorname{rank}D$ determines the source stratum. The kernel is the
Laplace transform that couples them.

The rank-one sector remains analytically controlled for arbitrary finite
$N$. For a general reduced source $D=uv^T$, the kernel reduces to the
ordinary HCIZ determinant with augmented spectra. Along the particularly
useful ray
\begin{equation}
D=rE_{11},
\end{equation}
the Haar marginal
$q_{11}=|U_{11}|^2\sim\operatorname{Beta}(1,N-1)$ gives the closed form
\begin{equation}
F_N(rE_{11})
=
{}_1F_1(1;N;r)
=
\frac{(N-1)!}{r^{N-1}}
\left(
e^r-\sum_{k=0}^{N-2}\frac{r^k}{k!}
\right).
\end{equation}
Thus the absence of a closed Haar-pushforward density for generic
$N\geq4$ does not imply the absence of analytic control. Rank-one sources
are exactly solvable at every finite $N$, while higher-rank sources may be
treated locally through Weingarten moments and at finite source by direct
Haar sampling. The numerical rank-two trajectory studied in
Section~\ref{sec:Haar-sampling} illustrates this complementary role of exact
and numerical methods.

The logarithm of the reduced kernel carries an additional geometric meaning.
On the reduced relative-coordinate slice,
\begin{equation}
K_N=\log F_N
\end{equation}
is a K\"ahler potential for the corresponding family of normalized
coherent states. Its first and second source derivatives are respectively
the mean and covariance of the exponentially tilted unistochastic ensemble,
\begin{equation}
\frac{\partial K_N}{\partial D_{ij}}
=
\langle q_{ij}\rangle_D,
\qquad
\frac{\partial^2 K_N}
{\partial D_{ij}\partial D_{kl}}
=
\operatorname{Cov}_D(q_{ij},q_{kl}).
\end{equation}
Consequently, the pullback Fubini--Study metric is determined directly by
unistochastic one- and two-point functions. This realizes, in the present
Haar-induced ensemble, the broader connection between fluctuation generating
functions and quantum-state geometry discussed in
Ref.~\cite{HetenyiLevay2023}. The additional structure here is the nonlinear
Gram-map pullback and the rank stratification of the physical source, which
make it possible to study not only the intrinsic metric but also the normal
and extrinsic geometry between lower- and higher-rank coherent-state
sectors. This turns the pushforward measure from a device for evaluating the
kernel into a geometric description of the protected coherent-state family.

The exact rank-one solution allows the embedding of the HCIZ sector into the
two-matrix coherent-state geometry to be studied at arbitrary finite $N$.
For the controlled family
\begin{equation}
\alpha=(a,0,\ldots,0),
\qquad
\beta=(u,v,0,\ldots,0),
\end{equation}
the surface $v=0$ lies in the rank-one locus, while $v$ supplies a
representative genuinely rank-two transverse direction. Writing
\begin{equation}
\mu_N(r)
=
\partial_r\log F_N(rE_{11}),
\qquad
s_N(r)
=
\partial_r^2\log F_N(rE_{11}),
\end{equation}
the induced normal metric is
\begin{equation}
h_N(r)
=
\frac{
N\mu_N(r)-1
}{
(N-1)^2\mu_N(r)
}.
\end{equation}
For every finite $N$ and nonzero physical background $r>0$, this quantity is
positive. Within the controlled family, the rank-one surface is therefore not
separated from the rank-two region by a finite-$N$ metric singularity; the
collapse of the representative normal direction occurs only at the rank-zero
apex. For $N\geq4$ the background has coincident eigenvalues, and some
rank-two directions outside the controlled family are null there.

At the same time, the HCIZ surface is not totally geodesic in the controlled
family, and at $N=3$ also in the full two-matrix geometry. On the controlled
surface $\mathcal H$ at $u=v=0$, its second fundamental form has the
nonvanishing component
\begin{equation}
B_{au}
=
-
\frac{
(N-1)\bar a\,s_N(r)
}{
N\mu_N(r)-1
}\,e,
\end{equation}
and the corresponding invariant norm on $\mathcal H$ satisfies
\begin{equation}
\|B\|^2
=
\frac{
2r\,s_N(r)^2
}{
\left(\mu_N+r s_N\right)
\mu_N^2
\left(N\mu_N-1\right)
}.
\end{equation}
The Gauss relation
\begin{equation}
R^{\mathcal M}_{a\bar a u\bar u}
-
R^{\mathcal H}_{a\bar a u\bar u}
=
\|B_{au}\|^2
\end{equation}
shows explicitly that part of the intrinsic curvature of the HCIZ family is
extrinsic: it arises from the bending of the rank-one state manifold inside
the larger two-matrix geometry. The rank-one sector is therefore an exact
analytic subfamily, but it is not geometrically closed under ambient
parallel transport.

The same embedding admits a controlled large-$N$ limit. Scaling
\begin{equation}
r=N\rho
\end{equation}
produces a boundary-to-interior saddle transition at $\rho=1$. The
finite-rank free energy is
\begin{equation}
\Phi(\rho)
=
\lim_{N\rightarrow\infty}
\frac{1}{N}
\log F_N(N\rho E_{11})
=
\begin{cases}
0,
&
\rho\leq1,
\\[1mm]
\rho-1-\log\rho,
&
\rho>1.
\end{cases}
\end{equation}
For $\rho<1$, the tilted matrix element remains of order $N^{-1}$; at the
critical point it scales as $N^{-1/2}$; and for $\rho>1$ it becomes
order one. This saddle structure is consistent with the known large-$N$
behavior of finite-rank spherical integrals and should not by itself be
interpreted as a new dynamical phase transition of the gauge theory.

What is specific to the present coherent-state problem is the way the same
saddle transition reorganizes the embedding geometry. In the controlled
slice,
\begin{equation}
N h_N(N\rho)
\longrightarrow
\min(\rho,1),
\qquad
\|B\|^2
\longrightarrow
2(1-\rho)_+.
\end{equation}
Below the threshold the rank-one family retains order-one extrinsic bending,
whereas above it the representative rank-two normal direction becomes
asymptotically geodesic at leading order. The corresponding intrinsic and
ambient mixed bisectional curvatures remain separated for $\rho<1$ but both
vanish on the interior-saddle side. In this controlled sense, the
effectively one-scalar coherent-state sector becomes geometrically more
autonomous at sufficiently strong scaled source, without implying a
dynamical decoupling or stability statement for individual states.

Several extensions are suggested by this picture. For more than two
commuting matrices, the same unistochastic measure appears while the
physical source fills higher-rank strata
\begin{equation}
\operatorname{rank}D\leq m.
\end{equation}
It would be useful to determine how the normal geometry generalizes from the
rank-one/rank-two interface studied here to the full nested sequence
\begin{equation}
\mathcal R_1\subset\mathcal R_2\subset\cdots
\subset\mathcal R_{\min(m,N-1)}.
\end{equation}
A second direction is to obtain better analytic control of the
Haar-pushforward measure for $N\geq4$, either through moment reconstruction,
large-deviation methods, or approximations adapted to fixed source rank.
The generalized unistochastic constructions of
\cite{nechita2025generalized} may also provide useful organizational tools
for this problem.

It would also be interesting to connect the rank-stratified source geometry
more directly with the principal-chiral-model and quenched Eguchi--Kawai
descriptions of multi-matrix BPS correlators
\cite{anempodistov2026solvable}. In those formulations the number of active
matrix directions controls the dimension of the corresponding reduced
momentum data. In the present variables, double differences of the source
are naturally Gram matrices of relative eigenvalue or momentum vectors, so
the source rank provides a finite-$N$ invariant with which to compare the
two descriptions. Establishing this correspondence beyond the controlled
families studied here may help connect the exact finite-$N$ geometry to
large-$N$ collective-coordinate dynamics.

The main conclusion is therefore not only that multi-matrix coherent-state
kernels admit a unistochastic reduction, but that this reduction exposes a
natural rank-stratified geometry of protected state space. Rank one is the
ordinary HCIZ sector; $N=3$ is the first value at which genuine rank-two
geometry appears; the rank-one family remains exactly solvable for all
finite $N$; and its embedding into a controlled slice of the two-matrix
state manifold is smooth but not totally geodesic. In a controlled large-$N$ limit, the same
finite-rank spherical-integral saddle that reorganizes the kernel also
reorganizes this embedding geometry. The resulting framework separates the
geometry of the integration domain from the geometry of the physical source
and provides a finite-$N$ route from exact BPS coherent-state overlaps to
their large-$N$ rank structure.

\section*{Acknowledgments}

The author used OpenAI’s ChatGPT (GPT-5.6 Sol, accessed September 2026) to assist with algebraic checks, numerical implementation, literature searches, and manuscript development and editing. The author reviewed the resulting calculations, references, and conclusions and takes responsibility for the content of the manuscript.  

\appendix

\section{Relation to the direct $U(3)$ parameterization}
\label{app:direct-U3}

The direct $U(3)$ calculation of
\cite{holguin2024multi} employed an Euler-angle parameterization of the
full unitary matrix. We show here explicitly how the exponent in that
parameterization reduces to the four moduli-squared variables used in
Section~\ref{sec:U3-kernel}, and hence to the same reduced source matrix
$D$ that underlies the rank decomposition of the main text.

To avoid confusion with the unistochastic coordinate $b$ used in the
main text, we denote by $\gamma$ the angle called $b$ in
\cite{holguin2024multi}, and define
\begin{equation}
\chi=\sigma+a .
\end{equation}
It is also convenient to introduce
\begin{equation}
\Delta
=
\frac12
\cos\theta\,
\sin(2\beta)\,
\sin(2\gamma)\,
\cos(2\chi).
\label{eq:app-Delta}
\end{equation}

To match the source-index convention used in the main text, we define the
unistochastic matrix below using the transpose convention
\begin{equation}
q_{ij}=|U_{ji}|^2.
\end{equation}
This transpose is a convention used only for matching the source indices of
Ref.~\cite{holguin2024multi}; it does not affect the doubly stochastic or
unistochastic properties of $q$. Writing $c_\eta=\cos\eta$ and
$s_\eta=\sin\eta$ for any angle $\eta$, the squared moduli of the unitary
matrix in the parameterization of Ref.~\cite{holguin2024multi} are
\begin{align}
q_{11}
&=
c_\beta^2 c_\theta^2 c_\gamma^2
+s_\beta^2 s_\gamma^2-\Delta,
&
q_{12}
&=
s_\beta^2 c_\theta^2 c_\gamma^2
+c_\beta^2 s_\gamma^2+\Delta,
\nonumber\\
q_{13}
&=
c_\gamma^2 s_\theta^2,
&
q_{21}
&=
s_\beta^2 c_\gamma^2
+c_\beta^2 c_\theta^2 s_\gamma^2+\Delta,
\nonumber\\
q_{22}
&=
c_\beta^2 c_\gamma^2
+s_\beta^2 c_\theta^2 s_\gamma^2-\Delta,
&
q_{23}
&=
s_\gamma^2 s_\theta^2,
\nonumber\\
q_{31}
&=
c_\beta^2 s_\theta^2,
&
q_{32}
&=
s_\beta^2 s_\theta^2,
\qquad
q_{33}=c_\theta^2 .
\label{eq:app-qij}
\end{align}

These quantities obey
\begin{equation}
\sum_j q_{ij}=1,
\qquad
\sum_i q_{ij}=1,
\end{equation}
as required by unitarity. Grouping the source-dependent terms in
Eq.~(A.4) of \cite{holguin2024multi} according to
\begin{equation}
C_{ij}=a_i\bar a_j+b_i\bar b_j
\end{equation}
gives simply
\begin{equation}
\operatorname{Tr}
\left(
U\bar\Lambda_XU^\dagger\Lambda_X+
U\bar\Lambda_YU^\dagger\Lambda_Y
\right)
=
\sum_{i,j=1}^3 C_{ij}q_{ij}.
\label{eq:app-Cq}
\end{equation}

Using double stochasticity to eliminate the last row and column gives
\begin{equation}
\sum_{i,j=1}^3 C_{ij}q_{ij}
=
R_3(C)
+xq_{11}+yq_{12}+zq_{21}+wq_{22},
\label{eq:app-reduced-exponent}
\end{equation}
where
\begin{equation}
R_3(C)
=
C_{13}+C_{23}+C_{31}+C_{32}-C_{33},
\end{equation}
and
\begin{equation}
\begin{aligned}
x&=C_{11}-C_{13}-C_{31}+C_{33},\\
y&=C_{12}-C_{13}-C_{32}+C_{33},\\
z&=C_{21}-C_{23}-C_{31}+C_{33},\\
w&=C_{22}-C_{23}-C_{32}+C_{33}.
\end{aligned}
\end{equation}
Equivalently,
\begin{equation}
D=
\begin{pmatrix}
x&y\\
z&w
\end{pmatrix},
\end{equation}
in agreement with Eq.~\eqref{eq:U3-D-matrix}.

For example, the only dependence on the combination
$\chi=\sigma+a$ is contained in $\Delta$. Its contribution to
Eq.~\eqref{eq:app-reduced-exponent} is
\begin{equation}
-\frac12
\left(
C_{11}-C_{12}-C_{21}+C_{22}
\right)
\cos\theta\,
\sin(2\beta)\,
\sin(2\gamma)\,
\cos(2\chi),
\end{equation}
which is precisely the sum of the two terms proportional to
$e^{\pm2i(\sigma+a)}$ in Eq.~(A.4) of
\cite{holguin2024multi}.

Thus the angular expression obtained in the direct calculation is
identically the reduced exponent used in the present work. The phase
degrees of freedom that are not independently resolved by the moduli
$q_{ij}$ lie along the fibers of the map
\begin{equation}
U(3)\longrightarrow\mathcal U_3,
\qquad
U\longmapsto |U|^2 .
\end{equation}
Passing to the Haar pushforward integrates over these fibers before the
source-dependent integral is performed. The direct Euler-angle calculation
and the unistochastic formulation therefore evaluate the same Haar average;
the latter simply uses variables adapted from the outset to the dependence
of the kernel on $|U_{ij}|^2$.

\section{Finite-$N$ geometry of the controlled rank-one embedding}
\label{app:finite-N-geometry}

This appendix collects the algebra underlying the finite-$N$ geometric
results of Section~\ref{sec:quantum-geometry}. We first derive the pullback
metric from the cumulant-generating function, then specialize to the
three-complex-dimensional family \eqref{eq:controlled-family}, construct the
normal line to the HCIZ locus, and verify the second fundamental form and
Gauss--Codazzi relations used in the main text.

\subsection{Pullback of the cumulant metric}
\label{app:pullback-metric}

On the reduced relative-coordinate slice, the K\"ahler potential is
\begin{equation}
K_N(D)=\log F_N(D),
\end{equation}
with
\begin{equation}
D_{ij}
=
\sum_A \xi_{iA}\bar\xi_{jA}.
\label{eq:appB-D-xi}
\end{equation}
We write
\begin{equation}
M_{ij}
=
\frac{\partial K_N}{\partial D_{ij}},
\qquad
\Gamma_{ij,kl}
=
\frac{\partial^2 K_N}
{\partial D_{ij}\partial D_{kl}}.
\label{eq:appB-M-Gamma}
\end{equation}
By the tilted-ensemble interpretation of Section~\ref{sec:cumulants},
\begin{equation}
M_{ij}=\langle q_{ij}\rangle_D,
\qquad
\Gamma_{ij,kl}
=
\operatorname{Cov}_D(q_{ij},q_{kl}).
\end{equation}

The derivatives of the reduced source are
\begin{equation}
\frac{\partial D_{np}}{\partial \xi_{iA}}
=
\delta_{ni}\bar\xi_{pA},
\qquad
\frac{\partial D_{np}}{\partial \bar\xi_{jB}}
=
\delta_{pj}\xi_{nB}.
\label{eq:appB-D-derivatives}
\end{equation}
Hence
\begin{equation}
\frac{\partial K_N}{\partial \xi_{iA}}
=
\sum_{n=1}^{N-1}
\bar\xi_{nA}M_{in}.
\label{eq:appB-first-xi-derivative}
\end{equation}
Differentiating once more gives
\begin{equation}
\begin{split}
g_{iA,\bar jB}
&=
\frac{\partial^2 K_N}
{\partial \xi_{iA}\partial\bar\xi_{jB}}
\\
&=
\delta_{AB}M_{ij}
+
\sum_{n,p=1}^{N-1}
\bar\xi_{nA}\xi_{pB}\,
\Gamma_{in,pj},
\end{split}
\label{eq:appB-pullback-metric}
\end{equation}
which is Eq.~\eqref{eq:general-pullback-metric}.

At the apex $\xi=0$, Haar symmetry gives $M_{ij}=1/N$, so for each scalar
label $A$ the corresponding block is
\begin{equation}
g^{(A)}(0)
=
\frac{1}{N}\,
\mathbf J_{N-1},
\label{eq:appB-apex-block}
\end{equation}
where $\mathbf J_{N-1}$ is the all-ones matrix. Each such block has one
nonzero eigenvalue $(N-1)/N$ and $N-2$ null eigenvalues. For $N=3$ and two
scalar matrices this reproduces Eq.~\eqref{eq:N3-vacuum-spectrum}. The origin
of this degeneracy is Eq.~\eqref{eq:apex-variations}: all first-order
variations of the state at the apex are proportional.

\subsection{The controlled family and its metric}
\label{app:controlled-metric-derivation}

Consider the family
\begin{equation}
\alpha=(a,0,\ldots,0),
\qquad
\beta=(u,v,0,\ldots,0),
\qquad
r=|a|^2,
\end{equation}
introduced in Eq.~\eqref{eq:controlled-family}. Along the background
$u=v=0$ the reduced source is
\begin{equation}
D=rE_{11}.
\end{equation}
Define
\begin{equation}
k_N(r)=\log F_N(rE_{11}),
\qquad
\mu_N=k_N',
\qquad
s_N=k_N'',
\qquad
t_N=k_N'''.
\label{eq:appB-kmst}
\end{equation}
The tilted measure depends only on $q_{11}$, so permutations of rows
$2,\ldots,N$ and columns $2,\ldots,N$ remain symmetries.

Since every row and column of $q$ sums to one,
\begin{equation}
\sum_{j=1}^{N}M_{1j}=1,
\qquad
\sum_{i=1}^{N}M_{i1}=1.
\end{equation}
With $M_{11}=\mu_N$, symmetry therefore gives
\begin{equation}
M_{1j}=M_{j1}
=
m_N
=
\frac{1-\mu_N}{N-1},
\qquad j>1.
\label{eq:appB-m}
\end{equation}
For $i,j>1$ all entries are equal, and another row sum yields
\begin{equation}
M_{ij}
=
d_N
=
\frac{N-2+\mu_N}{(N-1)^2}.
\label{eq:appB-d}
\end{equation}

The only covariance needed for the radial component is
\begin{equation}
\Gamma_{11,11}
=
\frac{d\mu_N}{dr}
=
s_N.
\end{equation}
Using Eq.~\eqref{eq:appB-pullback-metric}, the metric at $u=v=0$ in the
coordinate basis $(\partial_a,\partial_u,\partial_v)$ becomes
\begin{equation}
g
=
\begin{pmatrix}
A_N&0&0\\
0&\mu_N&m_N\\
0&m_N&d_N
\end{pmatrix},
\qquad
A_N=\mu_N+r s_N.
\label{eq:appB-controlled-metric}
\end{equation}
This reproduces Eqs.~\eqref{eq:controlled-metric} and
\eqref{eq:Amd-def}. The $(u,v)$ block has determinant
$\mu_Nd_N-m_N^2=\mu_Nh_N>0$ for $r>0$, so the metric restricted to the
controlled family is nondegenerate, even though for $N\geq4$ the full
pullback metric is degenerate at this background
(Section~\ref{sec:normal-geometry-general-N}).

On the HCIZ surface
\begin{equation}
\mathcal H:\qquad v=0,
\end{equation}
the reduced source has only one nonzero row and column after a suitable
rank-one factorization. More directly, within the controlled family,
\begin{equation}
D\big|_{\mathcal H}
=
\begin{pmatrix}
|a|^2+|u|^2&0\\
0&0
\end{pmatrix}
\oplus\mathbf 0_{N-3}.
\end{equation}
Hence the induced K\"ahler potential is
\begin{equation}
K_{\mathcal H}
=
k_N(R),
\qquad
R=|a|^2+|u|^2,
\label{eq:appB-KH}
\end{equation}
as used in Eq.~\eqref{eq:H-kahler-potential}.

\subsection{Normal line and second fundamental form}
\label{app:normal-B-derivation}

At $u=v=0$, the algebraic transverse vector $\partial_v$ is not orthogonal
to the tangent vector $\partial_u$, since
\begin{equation}
g_{v\bar u}=m_N.
\end{equation}
Subtracting its tangent projection gives
\begin{equation}
n
=
\partial_v-\frac{m_N}{\mu_N}\partial_u
=
\partial_v
-
\frac{1-\mu_N}{(N-1)\mu_N}\partial_u.
\label{eq:appB-normal-vector}
\end{equation}
Its norm is the Schur complement of the $\partial_u$ block:
\begin{equation}
\begin{split}
h_N
&=
d_N-\frac{m_N^2}{\mu_N}
\\
&=
\frac{N\mu_N-1}{(N-1)^2\mu_N}.
\end{split}
\label{eq:appB-normal-metric}
\end{equation}
This is Eq.~\eqref{eq:normal-metric}. Since
\begin{equation}
\mu_N'(r)=s_N(r)>0
\end{equation}
for the nondegenerate tilted beta distribution and
$\mu_N(0)=1/N$, one has $h_N(r)>0$ for $r>0$.

To compute the second fundamental form, it is enough to evaluate the
ambient K\"ahler connection along $\mathcal H$ at $u=0$. The derivatives of
the relevant metric components are
\begin{equation}
\partial_a g_{u\bar u}
=
\bar a\,s_N,
\qquad
\partial_a g_{u\bar v}
=
-\frac{\bar a\,s_N}{N-1}.
\label{eq:appB-metric-derivatives-B}
\end{equation}
The inverse of the $(u,v)$ block is
\begin{equation}
\begin{pmatrix}
\mu_N&m_N\\
m_N&d_N
\end{pmatrix}^{-1}
=
\frac{1}{\mu_N d_N-m_N^2}
\begin{pmatrix}
d_N&-m_N\\
-m_N&\mu_N
\end{pmatrix}.
\label{eq:appB-uv-inverse}
\end{equation}
Therefore
\begin{equation}
\Gamma^{u}_{au}
=
\frac{(N-1)\bar a\,s_N}{N\mu_N-1},
\qquad
\Gamma^{v}_{au}
=
-\frac{(N-1)\bar a\,s_N}{N\mu_N-1}.
\label{eq:appB-Christoffel-au}
\end{equation}

Let $e=[\partial_v]$ denote the quotient frame of the normal line
$T\mathcal M|_{\mathcal H}/T\mathcal H$. Modulo the tangent direction
$\partial_u$, Eq.~\eqref{eq:appB-Christoffel-au} gives
\begin{equation}
B_{au}=B_{ua}
=
-
\frac{(N-1)\bar a\,s_N}{N\mu_N-1}\,e.
\label{eq:appB-Bau}
\end{equation}
Direct differentiation shows that the remaining independent components
have no normal quotient part:
\begin{equation}
B_{aa}=B_{uu}=0.
\label{eq:appB-B-diagonal}
\end{equation}
Thus the controlled HCIZ embedding is not totally geodesic for any finite
$N$ and $r>0$.

The quotient frame $e$ has norm $h_N$, so
\begin{equation}
\begin{split}
\|B_{au}\|^2
&=
\left|
\frac{(N-1)\bar a\,s_N}{N\mu_N-1}
\right|^2 h_N
\\
&=
\frac{r s_N^2}
{\mu_N(N\mu_N-1)}.
\end{split}
\label{eq:appB-Bau-norm}
\end{equation}
At $u=0$ the tangent metric is diagonal,
\begin{equation}
g_{\mathcal H}
=
\operatorname{diag}(A_N,\mu_N),
\end{equation}
and the symmetry $B_{au}=B_{ua}$ therefore gives
\begin{equation}
\begin{split}
\|B\|^2
&=
2\,g^{a\bar a}g^{u\bar u}\|B_{au}\|^2
\\
&=
\frac{
2r s_N^2
}{
(\mu_N+r s_N)\mu_N^2(N\mu_N-1)
}.
\end{split}
\label{eq:appB-B-full}
\end{equation}

The small-$r$ behavior follows directly from the first two cumulants of
$q_{11}\sim\operatorname{Beta}(1,N-1)$ at the origin:
\begin{equation}
\mu_N(0)=\frac1N,
\qquad
s_N(0)
=
\frac{N-1}{N^2(N+1)}.
\label{eq:appB-origin-cumulants}
\end{equation}
Thus
\begin{equation}
\mu_N(r)
=
\frac1N
+
\frac{N-1}{N^2(N+1)}\,r
+
O(r^2),
\end{equation}
which gives
\begin{equation}
h_N(r)
=
\frac{r}{N^2-1}
+
O(r^2),
\label{eq:appB-h-small}
\end{equation}
and
\begin{equation}
\|B\|^2
=
\frac{2(N-1)}{N+1}
+
O(r).
\label{eq:appB-B-small}
\end{equation}

For fixed $N$ and $r\to\infty$, the exact expression
\eqref{eq:rank-one-elementary} gives
\begin{equation}
k_N(r)
=
r-(N-1)\log r+\log (N-1)!
+O\!\left(r^{N-2}e^{-r}\right).
\end{equation}
Consequently, up to corrections of order $r^{N-2}e^{-r}$,
\begin{equation}
\mu_N
=
1-\frac{N-1}{r},
\qquad
s_N
=
\frac{N-1}{r^2},
\qquad
\mu_N+r s_N=1,
\end{equation}
and therefore, up to exponentially small corrections,
\begin{equation}
h_N(r)
=
\frac{1}{N-1}
\left(
1-\frac{1}{r-N+1}
\right),
\qquad
\|B\|^2
=
\frac{2(N-1)}{(r-N)(r-N+1)^2}.
\label{eq:appB-fixed-N-large-r}
\end{equation}
At leading order these reduce to $h_N\to1/(N-1)$ and
$\|B\|^2\sim2(N-1)/r^3$, Eqs.~\eqref{eq:h-large-r-fixed-N} and
\eqref{eq:B-large-r-fixed-N}.

\paragraph{Full $N=3$ geometry.}
\label{app:full-space-N3}
For $N=3$ the controlled family is the hypersurface $\alpha_2=0$ in the full
space of relative coordinates $(\alpha_1,\alpha_2,\beta_1,\beta_2)$, and the
rank-one locus is the hypersurface $\alpha_1\beta_2-\alpha_2\beta_1=0$. At
$\alpha=(a,0)$, $\beta=0$ the tangent space of the latter is spanned by
$\partial_{\alpha_1}$, $\partial_{\alpha_2}$, and $\partial_{\beta_1}$. Since
$\xi_{iY}=0$ there, Eq.~\eqref{eq:appB-pullback-metric} gives
$g_{\alpha_i\bar\beta_j}=0$ and $\partial_{\alpha_1}g_{\beta_1\bar\alpha_j}=0$,
so the $\alpha$- and $\beta$-blocks decouple. The normal line is therefore
spanned by the vector \eqref{eq:appB-normal-vector} with $u=\beta_1$ and
$v=\beta_2$, and $\Gamma^{\beta_2}_{\alpha_1\beta_1}$ is computed entirely
within the $\beta$-block, where it reduces to $\Gamma^v_{au}$ in
Eq.~\eqref{eq:appB-Christoffel-au}. Hence Eq.~\eqref{eq:appB-Bau} is also the
$(\alpha_1,\beta_1)$ component of the second fundamental form of the rank-one
hypersurface in the full $N=3$ two-matrix geometry. The $\beta$-block is the
$(u,v)$ block above, and a direct numerical evaluation shows that the
$\alpha$-block, $M_{ij}+r\,\Gamma_{i1,1j}$, likewise has no null directions for
$r>0$.

\subsection{Curvature and the Gauss equation}
\label{app:curvature-derivation}

We use the curvature convention of Eq.~\eqref{eq:curvature-convention},
\begin{equation}
R_{i\bar j k\bar l}
=
-\partial_i\partial_{\bar j}g_{k\bar l}
+
g^{p\bar q}
(\partial_i g_{k\bar q})
(\partial_{\bar j}g_{p\bar l}).
\label{eq:appB-curvature-convention}
\end{equation}
On $\mathcal H$, the potential is $k_N(R)$ with
$R=|a|^2+|u|^2$. At $u=0$,
\begin{equation}
g^{\mathcal H}_{a\bar a}
=
\mu_N+r s_N,
\qquad
g^{\mathcal H}_{u\bar u}
=
\mu_N,
\qquad
g^{\mathcal H}_{a\bar u}=0.
\end{equation}
The derivatives needed for the mixed curvature are
\begin{equation}
\partial_a g^{\mathcal H}_{u\bar u}
=
\bar a\,s_N,
\end{equation}
and
\begin{equation}
\partial_a\partial_{\bar a}
g^{\mathcal H}_{u\bar u}
=
s_N+r t_N.
\end{equation}
Equation~\eqref{eq:appB-curvature-convention} then gives
\begin{equation}
R^{\mathcal H}_{a\bar a u\bar u}
=
-s_N-r t_N
+
\frac{r s_N^2}{\mu_N}.
\label{eq:appB-RH}
\end{equation}

A direct evaluation of the corresponding ambient component using the full
$(a,u,v)$ metric gives
\begin{equation}
R^{\mathcal M}_{a\bar a u\bar u}
=
-s_N-r t_N
+
\frac{N r s_N^2}{N\mu_N-1}.
\label{eq:appB-RM}
\end{equation}
Subtracting Eqs.~\eqref{eq:appB-RH} and \eqref{eq:appB-RM} yields
\begin{equation}
\begin{split}
R^{\mathcal M}_{a\bar a u\bar u}
-
R^{\mathcal H}_{a\bar a u\bar u}
&=
r s_N^2
\left[
\frac{N}{N\mu_N-1}
-\frac{1}{\mu_N}
\right]
\\
&=
\frac{r s_N^2}
{\mu_N(N\mu_N-1)}
\\
&=
\|B_{au}\|^2,
\end{split}
\label{eq:appB-Gauss-mixed}
\end{equation}
which verifies the mixed Gauss equation in the convention used in the main
text.

At $u=v=0$, because $B_{aa}=B_{uu}=0$ and the tangent metric is diagonal,
the only nonzero contributions to the scalar trace of the Gauss equation are
the two mixed contractions. Thus
\begin{equation}
S_{\mathcal M}^{\parallel}
-
S_{\mathcal H}
=
2\,g^{a\bar a}g^{u\bar u}\|B_{au}\|^2
=
\|B\|^2,
\label{eq:appB-Gauss-scalar}
\end{equation}
which gives Eq.~\eqref{eq:scalar-gauss}.

\subsection{Normal-bundle curvature and Codazzi relation}
\label{app:normal-curvature-codazzi}

The normal line has Hermitian metric
\begin{equation}
h_N(r)
=
\frac{N\mu_N-1}{(N-1)^2\mu_N}.
\end{equation}
For any radial function $f(r)$ with $r=|a|^2$,
\begin{equation}
\partial_a\partial_{\bar a}f(r)
=
f'(r)+r f''(r).
\label{eq:appB-radial-ddbar}
\end{equation}
Using
\begin{equation}
\frac{d}{dr}\log h_N
=
\frac{N s_N}{N\mu_N-1}
-
\frac{s_N}{\mu_N},
\end{equation}
and differentiating once more, the Chern curvature
\begin{equation}
\Theta^\perp
=
-\partial\bar\partial\log h_N
\end{equation}
has radial component
\begin{equation}
\Theta^\perp_{a\bar a}
=
-
\frac{s_N+r t_N}
{\mu_N(N\mu_N-1)}
+
\frac{
r s_N^2(2N\mu_N-1)
}{
\mu_N^2(N\mu_N-1)^2
}.
\label{eq:appB-normal-curvature}
\end{equation}

For the Codazzi relation, write
\begin{equation}
B_{au}=b_N(r,\bar a)\,e,
\qquad
b_N
=
-\frac{(N-1)\bar a\,s_N}{N\mu_N-1}.
\end{equation}
In the holomorphic quotient frame $e$, the $(0,1)$ part of the Chern
connection is simply $\bar\partial$, and therefore
\begin{equation}
-\nabla^\perp_{\bar a}B_{au}
=
-\partial_{\bar a}b_N\,e.
\end{equation}
Since
\begin{equation}
\partial_{\bar a}
\left[
\bar a\,
\frac{s_N}{N\mu_N-1}
\right]
=
\frac{
(s_N+r t_N)(N\mu_N-1)-N r s_N^2
}{
(N\mu_N-1)^2
},
\end{equation}
one obtains
\begin{equation}
-\left(\nabla^\perp_{\bar a}B\right)_{au}
=
\mathcal C_N(r)e,
\end{equation}
where
\begin{equation}
\mathcal C_N(r)
=
(N-1)
\frac{
(s_N+r t_N)(N\mu_N-1)-N r s_N^2
}{
(N\mu_N-1)^2
}.
\label{eq:appB-Codazzi}
\end{equation}
This is the coefficient appearing in
Eq.~\eqref{eq:codazzi-relation}.

\subsection{$N=3$ specialization and numerical check}
\label{app:N3-geometry-check}

For $N=3$,
\begin{equation}
F_3(rE_{11})
=
\frac{2(e^r-1-r)}{r^2},
\end{equation}
and hence
\begin{equation}
\mu_3(r)
=
\frac{e^r-1}{e^r-1-r}
-\frac{2}{r}
=
\frac{e^r(r-2)+r+2}
{r(e^r-1-r)}.
\label{eq:appB-mu3}
\end{equation}
Its expansion at the origin is
\begin{equation}
\mu_3(r)
=
\frac13
+\frac{r}{18}
+\frac{r^2}{270}
-\frac{r^3}{3240}
-\frac{r^4}{13608}
+O(r^5).
\label{eq:appB-mu3-series}
\end{equation}
The normal metric becomes
\begin{equation}
h_3(r)
=
\frac{3\mu_3(r)-1}{4\mu_3(r)},
\end{equation}
with
\begin{equation}
h_3(r)
=
\frac{r}{8}
-\frac{r^2}{80}
+\frac{r^4}{11200}
+O(r^6).
\label{eq:appB-h3-series}
\end{equation}
The invariant norm of the second fundamental form has the expansion
\begin{equation}
\|B\|^2
=
1
-\frac{7r}{15}
+\frac{7r^2}{75}
-\frac{19r^3}{2250}
+O(r^4).
\label{eq:appB-B3-series}
\end{equation}

The radial metric may be written in closed form as
\begin{equation}
g^{(3)}_{a\bar a}
=
\mu_3+r s_3
=
\frac{
e^{2r}-2e^r+1-r^2e^r
}{
(e^r-1-r)^2
},
\end{equation}
whose small-$r$ expansion is
\begin{equation}
g^{(3)}_{a\bar a}
=
\frac13
+\frac{r}{9}
+\frac{r^2}{90}
-\frac{r^3}{810}
+O(r^4).
\end{equation}

As a finite-source check, at $r=1$,
\begin{equation}
\begin{aligned}
\mu_3&=0.3922111912,
&
s_3&=0.0617479992,
&
t_3&=0.0047010626,
\\
A_3&=0.4539591903,
&
h_3&=0.1125883054.
\end{aligned}
\label{eq:appB-N3-values}
\end{equation}
The component and total extrinsic norms are
\begin{equation}
\|B_{au}\|^2
=
0.0550367198,
\qquad
\|B\|^2
=
0.6182238335.
\end{equation}
The intrinsic and ambient mixed curvature components are
\begin{equation}
R^{\mathcal H}_{a\bar a u\bar u}
=
-0.0567277293,
\qquad
R^{\mathcal M}_{a\bar a u\bar u}
=
-0.0016910094,
\end{equation}
and their difference,
\begin{equation}
0.0550367198,
\end{equation}
agrees with $\|B_{au}\|^2$ as required by
Eq.~\eqref{eq:appB-Gauss-mixed}. Dividing by
$g_{a\bar a}g_{u\bar u}=A_3\mu_3$ gives the mixed bisectional curvatures
\begin{equation}
\mathcal K_{\mathcal H}
=
-0.3186094152,
\qquad
\mathcal K_{\mathcal M}
=
-0.0094974985.
\end{equation}
These values provide a direct numerical check of the finite-$N$ curvature
identities used in Section~\ref{sec:gauss-codazzi}.

\section{Large-$N$ asymptotics of the rank-one ray}
\label{app:large-N-asymptotics}

This appendix gives the saddle-point derivations underlying
Section~\ref{sec:large-N-geometry}. We begin from the exact finite-$N$
rank-one integral
\begin{equation}
F_N(rE_{11})
=
(N-1)\int_0^1 dq\,
e^{rq}(1-q)^{N-2},
\label{eq:appC-start}
\end{equation}
and study the scaling
\begin{equation}
r=N\rho,
\qquad
\rho>0.
\label{eq:appC-scaling}
\end{equation}
The asymptotic analysis has three distinct regimes: a boundary saddle for
$0<\rho<1$, an interior saddle for $\rho>1$, and an $N^{-1/2}$ crossover
window around $\rho=1$. We then insert the resulting cumulants into the
finite-$N$ formulas of Appendix~\ref{app:finite-N-geometry}.

Throughout this appendix, asymptotic statements at fixed $\rho<1$ or fixed
$\rho>1$ are understood to hold uniformly on compact subsets bounded away
from the transition point $\rho=1$.

\subsection{Boundary saddle for $0<\rho<1$}
\label{app:boundary-saddle}

With $r=N\rho$, Eq.~\eqref{eq:appC-start} becomes
\begin{equation}
F_N(N\rho E_{11})
=
(N-1)\int_0^1dq\,
\exp\left[
N\rho q+(N-2)\log(1-q)
\right].
\label{eq:appC-beta-scaled}
\end{equation}
It is convenient to write
\begin{equation}
\phi_\rho(q)
=
\rho q+\log(1-q).
\end{equation}
For $0<\rho<1$,
\begin{equation}
\phi_\rho'(q)
=
\rho-\frac{1}{1-q}<0
\end{equation}
throughout the integration interval, so the maximum occurs at the endpoint
$q=0$.

The correct boundary scale is
\begin{equation}
q=\frac{x}{N}.
\label{eq:appC-boundary-scale}
\end{equation}
Substituting into Eq.~\eqref{eq:appC-beta-scaled} gives
\begin{equation}
\begin{split}
F_N(N\rho E_{11})
&=
\frac{N-1}{N}
\int_0^N dx\,
\exp\Bigg[
\rho x
+
(N-2)\log\left(1-\frac{x}{N}\right)
\Bigg]
\\
&=
\frac{N-1}{N}
\int_0^N dx\,
e^{-(1-\rho)x}
\exp\left[
\frac{2x-\frac12x^2}{N}
+O\!\left(\frac{x^3+x^2}{N^2}\right)
\right].
\end{split}
\label{eq:appC-boundary-expanded}
\end{equation}
For fixed $\rho<1$, the exponential
$e^{-(1-\rho)x}$ localizes the integral at $x=O(1)$, allowing the upper
limit to be extended to infinity with exponentially small error. Thus
\begin{equation}
F_N(N\rho E_{11})
=
\frac{1}{1-\rho}
\left[
1+O(N^{-1})
\right].
\label{eq:appC-F-below}
\end{equation}
In particular,
\begin{equation}
\frac{1}{N}\log F_N(N\rho E_{11})
\longrightarrow0.
\end{equation}

The same scaling determines the tilted distribution of $q_{11}$. To leading
order the variable
\begin{equation}
x=Nq_{11}
\end{equation}
has the exponential density
\begin{equation}
p_\rho(x)
=
(1-\rho)e^{-(1-\rho)x},
\qquad
x\geq0.
\label{eq:appC-boundary-exp-density}
\end{equation}
Hence
\begin{equation}
\langle q_{11}\rangle
=
\frac{1}{N(1-\rho)}
+O(N^{-2}),
\end{equation}
and
\begin{equation}
\operatorname{Var}(q_{11})
=
\frac{1}{N^2(1-\rho)^2}
+O(N^{-3}).
\end{equation}
Equivalently, in the notation of Eq.~\eqref{eq:mu-s-def},
\begin{equation}
\mu_N(N\rho)
=
\frac{1}{N(1-\rho)}
+O(N^{-2}),
\label{eq:appC-mu-below}
\end{equation}
\begin{equation}
s_N(N\rho)
=
\frac{1}{N^2(1-\rho)^2}
+O(N^{-3}),
\label{eq:appC-s-below}
\end{equation}
and differentiation with respect to $r=N\rho$ gives
\begin{equation}
t_N(N\rho)
=
\frac{2}{N^3(1-\rho)^3}
+O(N^{-4}).
\label{eq:appC-t-below}
\end{equation}

\subsection{Interior saddle for $\rho>1$}
\label{app:interior-saddle}

For $\rho>1$, the stationary-point equation
\begin{equation}
\phi_\rho'(q_\star)=0
\end{equation}
has the solution
\begin{equation}
q_\star
=
1-\frac{1}{\rho},
\label{eq:appC-qstar}
\end{equation}
which lies in the interior of the integration interval. At the saddle,
\begin{equation}
\phi_\rho(q_\star)
=
\rho-1-\log\rho,
\label{eq:appC-phistar}
\end{equation}
and
\begin{equation}
\phi_\rho''(q_\star)
=
-\rho^2.
\end{equation}

To keep the prefactor correct, rewrite
Eq.~\eqref{eq:appC-beta-scaled} as
\begin{equation}
F_N(N\rho E_{11})
=
(N-1)
\int_0^1dq\,
\frac{e^{N\phi_\rho(q)}}{(1-q)^2}.
\label{eq:appC-interior-form}
\end{equation}
The smooth amplitude is
\begin{equation}
g(q)=\frac{1}{(1-q)^2},
\qquad
g(q_\star)=\rho^2.
\end{equation}
Standard Laplace asymptotics therefore give
\begin{equation}
\begin{split}
F_N(N\rho E_{11})
&=
(N-1)\rho^2
\sqrt{\frac{2\pi}{N\rho^2}}\,
\exp\left[
N(\rho-1-\log\rho)
\right]
\left[
1+O(N^{-1})
\right]
\\
&=
\rho\sqrt{2\pi N}\,
\exp\left[
N(\rho-1-\log\rho)
\right]
\left[
1+O(N^{-1})
\right].
\end{split}
\label{eq:appC-F-above}
\end{equation}
Consequently,
\begin{equation}
\lim_{N\to\infty}
\frac{1}{N}\log F_N(N\rho E_{11})
=
\rho-1-\log\rho.
\end{equation}

Combining this with the boundary result yields
\begin{equation}
\Phi(\rho)
=
\lim_{N\to\infty}
\frac1N\log F_N(N\rho E_{11})
=
\begin{cases}
0,
&
0<\rho\leq1,
\\[1mm]
\rho-1-\log\rho,
&
\rho>1.
\end{cases}
\label{eq:appC-free-energy}
\end{equation}
The function $\Phi$ and its first derivative are continuous at $\rho=1$,
whereas
\begin{equation}
\Phi''(\rho)
=
\begin{cases}
0,&\rho<1,\\
\rho^{-2},&\rho>1,
\end{cases}
\end{equation}
jumps at the transition.

The first derivative of the logarithm may be obtained either by
differentiating Eq.~\eqref{eq:appC-F-above} or by expanding the tilted
measure about $q_\star$. Since
\begin{equation}
\partial_r
=
\frac1N\partial_\rho,
\end{equation}
one finds
\begin{equation}
\mu_N(N\rho)
=
1-\frac1\rho
+
O(N^{-1}),
\label{eq:appC-mu-above}
\end{equation}
and
\begin{equation}
s_N(N\rho)
=
\frac{1}{N\rho^2}
+
O(N^{-2}).
\label{eq:appC-s-above}
\end{equation}
One further derivative gives
\begin{equation}
t_N(N\rho)
=
-\frac{2}{N^2\rho^3}
+
O(N^{-3}).
\label{eq:appC-t-above}
\end{equation}
Thus the tilted matrix element is centered at the order-one saddle value
$q_\star$, with fluctuations of width $N^{-1/2}$.

\subsection{Critical window around $\rho=1$}
\label{app:critical-window}

The fixed-$\rho$ expansions above cease to be uniform when
$|\rho-1|=O(N^{-1/2})$. Introduce the crossover variable
\begin{equation}
\rho
=
1+\frac{\lambda}{\sqrt N},
\qquad
\lambda=O(1),
\label{eq:appC-critical-rho}
\end{equation}
and scale
\begin{equation}
q=\frac{x}{\sqrt N}.
\label{eq:appC-critical-q}
\end{equation}
Then
\begin{equation}
\begin{split}
&N\rho q+(N-2)\log(1-q)
\\
&\qquad
=
\lambda x-\frac{x^2}{2}
+
\frac{1}{\sqrt N}
\left(
2x-\frac{x^3}{3}
\right)
+
O(N^{-1}).
\end{split}
\label{eq:appC-critical-exponent}
\end{equation}
Since
\begin{equation}
(N-1)dq
=
\sqrt N\left(1-\frac1N\right)dx,
\end{equation}
the leading crossover form is
\begin{equation}
F_N(N\rho E_{11})
=
\sqrt N\,\Psi(\lambda)
\left[
1+O(N^{-1/2})
\right],
\label{eq:appC-critical-F}
\end{equation}
where
\begin{equation}
\Psi(\lambda)
=
\int_0^\infty
e^{\lambda x-x^2/2}\,dx.
\label{eq:appC-Psi}
\end{equation}
Completing the square gives the equivalent closed form
\begin{equation}
\Psi(\lambda)
=
\sqrt{\frac{\pi}{2}}\,
e^{\lambda^2/2}
\operatorname{erfc}
\left(
-\frac{\lambda}{\sqrt2}
\right).
\label{eq:appC-Psi-erfc}
\end{equation}

At the transition $\lambda=0$,
\begin{equation}
\Psi(0)=\sqrt{\frac{\pi}{2}},
\end{equation}
and therefore
\begin{equation}
F_N(NE_{11})
\sim
\sqrt{\frac{\pi N}{2}}.
\label{eq:appC-F-critical}
\end{equation}
The typical matrix element is now of order $N^{-1/2}$ rather than
$N^{-1}$ or order one.

For later use define
\begin{equation}
m(\lambda)
=
\frac{d}{d\lambda}\log\Psi(\lambda),
\qquad
v(\lambda)
=
\frac{d^2}{d\lambda^2}\log\Psi(\lambda).
\label{eq:appC-mv}
\end{equation}
Since
\begin{equation}
\Psi'(\lambda)
=
\lambda\Psi(\lambda)+1,
\end{equation}
one may also write
\begin{equation}
m(\lambda)
=
\lambda+\frac{1}{\Psi(\lambda)}
\end{equation}
and
\begin{equation}
v(\lambda)
=
1+\lambda m(\lambda)-m(\lambda)^2.
\label{eq:appC-v-identity}
\end{equation}
Because
\begin{equation}
\partial_r
=
\frac{1}{\sqrt N}\partial_\lambda
\end{equation}
inside the critical window, Eq.~\eqref{eq:appC-critical-F} gives
\begin{equation}
\mu_N
=
\frac{m(\lambda)}{\sqrt N}
+
O(N^{-1}),
\label{eq:appC-critical-mu}
\end{equation}
and
\begin{equation}
s_N
=
\frac{v(\lambda)}{N}
+
O(N^{-3/2}).
\label{eq:appC-critical-s}
\end{equation}
A further derivative yields
\begin{equation}
t_N
=
\frac{v'(\lambda)}{N^{3/2}}
+
O(N^{-2}).
\label{eq:appC-critical-t}
\end{equation}

These formulas interpolate between the boundary and interior regimes. In
particular, the three characteristic scales of the tilted matrix element are
\begin{equation}
q_{11}
\sim
\begin{cases}
N^{-1},
&
0<\rho<1,
\\
N^{-1/2},
&
\rho=1,
\\
O(1),
&
\rho>1.
\end{cases}
\label{eq:appC-q-three-scales}
\end{equation}

\subsection{Asymptotics of the normal metric}
\label{app:large-N-h}

We now substitute the preceding cumulants into the exact finite-$N$ normal
metric
\begin{equation}
h_N(r)
=
\frac{N\mu_N(r)-1}
{(N-1)^2\mu_N(r)}.
\label{eq:appC-h-exact}
\end{equation}

For fixed $0<\rho<1$, let $a_\rho=1-\rho$. Using
Eq.~\eqref{eq:appC-mu-below},
\begin{equation}
N\mu_N-1
=
\frac{1}{a_\rho}-1+O(N^{-1})
=
\frac{\rho}{a_\rho}
+O(N^{-1}),
\end{equation}
whereas
\begin{equation}
(N-1)^2\mu_N
=
\frac{N}{a_\rho}
\left[
1+O(N^{-1})
\right].
\end{equation}
Hence
\begin{equation}
h_N(N\rho)
=
\frac{\rho}{N}
+
O(N^{-2}),
\qquad
0<\rho<1.
\label{eq:appC-h-below}
\end{equation}

For fixed $\rho>1$, Eq.~\eqref{eq:appC-mu-above} gives
\begin{equation}
N\mu_N-1
=
N\left(1-\frac1\rho\right)+O(1)
\end{equation}
and
\begin{equation}
(N-1)^2\mu_N
=
N^2\left(1-\frac1\rho\right)+O(N),
\end{equation}
so
\begin{equation}
h_N(N\rho)
=
\frac1N+O(N^{-2}),
\qquad
\rho>1.
\label{eq:appC-h-above}
\end{equation}
The two regimes combine into
\begin{equation}
N h_N(N\rho)
\longrightarrow
\min(\rho,1).
\label{eq:appC-h-limit}
\end{equation}

Inside the critical window it is convenient to rewrite
Eq.~\eqref{eq:appC-h-exact} exactly as
\begin{equation}
N h_N
=
\left(\frac{N}{N-1}\right)^2
\left(
1-\frac{1}{N\mu_N}
\right).
\end{equation}
By Eq.~\eqref{eq:appC-critical-mu},
$N\mu_N=\sqrt N\,m(\lambda)+O(1)$, and therefore
\begin{equation}
N h_N
=
1-\frac{1}{\sqrt N\,m(\lambda)}
+O(N^{-1}).
\label{eq:appC-h-critical}
\end{equation}
As $\lambda\to-\infty$, $m(\lambda)=-1/\lambda+O(\lambda^{-3})$, so
$1/(\sqrt N\,m)\to-\lambda/\sqrt N=1-\rho$ and $Nh_N\to\rho$. As
$\lambda\to+\infty$, $m(\lambda)-\lambda$ is exponentially small and
$Nh_N\to1$. Thus finite $N$ smooths the cusp of the limiting function
$\min(\rho,1)$ over a window of width $N^{-1/2}$.

\subsection{Asymptotics of the second fundamental form}
\label{app:large-N-B}

The invariant squared norm of the second fundamental form is
\begin{equation}
\|B\|^2
=
\frac{
2r\,s_N^2
}{
(\mu_N+r s_N)
\mu_N^2
(N\mu_N-1)
}.
\label{eq:appC-B-exact}
\end{equation}

For fixed $0<\rho<1$, Eqs.~\eqref{eq:appC-mu-below} and
\eqref{eq:appC-s-below} imply
\begin{equation}
\mu_N+r s_N
=
\frac{1}{N(1-\rho)^2}
+
O(N^{-2}).
\end{equation}
Substitution into Eq.~\eqref{eq:appC-B-exact} gives
\begin{equation}
\|B\|^2
=
2(1-\rho)
+
O(N^{-1}),
\qquad
0<\rho<1.
\label{eq:appC-B-below}
\end{equation}

For fixed $\rho>1$,
\begin{equation}
\mu_N+r s_N
=
1+O(N^{-1}),
\end{equation}
and
\begin{equation}
N\mu_N-1
=
N\frac{\rho-1}{\rho}+O(1).
\end{equation}
Using Eq.~\eqref{eq:appC-s-above} then yields
\begin{equation}
\|B\|^2
=
\frac{2}{N^2(\rho-1)^3}
+
O(N^{-3}),
\qquad
\rho>1.
\label{eq:appC-B-above}
\end{equation}
Thus
\begin{equation}
\|B\|^2
\longrightarrow
2(1-\rho)_+,
\label{eq:appC-B-limit}
\end{equation}
where $(x)_+=\max(x,0)$.

In the critical window, using
\begin{equation}
r=N+O(\sqrt N),
\qquad
\mu_N=\frac{m}{\sqrt N}+O(N^{-1}),
\qquad
s_N=\frac{v}{N}+O(N^{-3/2}),
\end{equation}
one obtains
\begin{equation}
\mu_N+r s_N
=
v(\lambda)+O(N^{-1/2}),
\end{equation}
and hence
\begin{equation}
\|B\|^2
=
\frac{
2v(\lambda)
}{
m(\lambda)^3
}
N^{-1/2}
+
O(N^{-1}).
\label{eq:appC-B-critical}
\end{equation}
The extrinsic curvature therefore vanishes continuously through the
critical window, even though it remains order one at fixed
$\rho<1$ before the limit $\rho\to1^-$ is taken.

\subsection{Mixed bisectional curvatures}
\label{app:large-N-curvatures}

The intrinsic and ambient mixed curvature components derived in
Appendix~\ref{app:curvature-derivation} are
\begin{equation}
R^{\mathcal H}_{a\bar a u\bar u}
=
-s_N-r t_N
+
\frac{r s_N^2}{\mu_N},
\label{eq:appC-RH}
\end{equation}
and
\begin{equation}
R^{\mathcal M}_{a\bar a u\bar u}
=
-s_N-r t_N
+
\frac{N r s_N^2}{N\mu_N-1}.
\label{eq:appC-RM}
\end{equation}
The tangent metric factors at $u=0$ are
\begin{equation}
g_{a\bar a}
=
A_N
=
\mu_N+r s_N,
\qquad
g_{u\bar u}
=
\mu_N.
\end{equation}
Define the normalized mixed bisectional curvatures
\begin{equation}
\mathcal K_{\mathcal H}
=
\frac{
R^{\mathcal H}_{a\bar a u\bar u}
}{
A_N\mu_N
},
\qquad
\mathcal K_{\mathcal M}
=
\frac{
R^{\mathcal M}_{a\bar a u\bar u}
}{
A_N\mu_N
}.
\label{eq:appC-K-def}
\end{equation}

For $0<\rho<1$, write $a_\rho=1-\rho$. From
Eqs.~\eqref{eq:appC-mu-below}--\eqref{eq:appC-t-below},
\begin{equation}
A_N
=
\frac{1}{Na_\rho^2}
+O(N^{-2}),
\end{equation}
and therefore
\begin{equation}
A_N\mu_N
=
\frac{1}{N^2a_\rho^3}
+O(N^{-3}).
\end{equation}
The intrinsic numerator is
\begin{equation}
R^{\mathcal H}_{a\bar a u\bar u}
=
-\frac{1}{N^2a_\rho^3}
+O(N^{-3}),
\end{equation}
while the ambient numerator is
\begin{equation}
R^{\mathcal M}_{a\bar a u\bar u}
=
-\frac{\rho}{N^2a_\rho^3}
+O(N^{-3}).
\end{equation}
Consequently,
\begin{equation}
\mathcal K_{\mathcal H}
\longrightarrow
-1,
\qquad
\mathcal K_{\mathcal M}
\longrightarrow
-\rho,
\qquad
0<\rho<1.
\label{eq:appC-K-below}
\end{equation}
Their difference is
\begin{equation}
\mathcal K_{\mathcal M}
-
\mathcal K_{\mathcal H}
\longrightarrow
1-\rho,
\label{eq:appC-K-gap}
\end{equation}
which is the normalized Gauss contribution of the second fundamental form.

For fixed $\rho>1$, one instead has
\begin{equation}
A_N
=
1+O(N^{-1}),
\qquad
\mu_N
=
\frac{\rho-1}{\rho}+O(N^{-1}),
\end{equation}
and both curvature numerators are of order $N^{-1}$. Using
Eqs.~\eqref{eq:appC-mu-above}--\eqref{eq:appC-t-above},
$-s_N-r t_N=1/(N\rho^2)+O(N^{-2})$ and
$r s_N^2/\mu_N=1/\bigl(N\rho^2(\rho-1)\bigr)+O(N^{-2})$, so
\begin{equation}
R^{\mathcal H}_{a\bar a u\bar u}
=
\frac{1}{N\rho(\rho-1)}
+O(N^{-2}).
\end{equation}
Hence, for $\rho>1$,
\begin{equation}
\begin{split}
\mathcal K_{\mathcal H}
&=
\frac{1}{N(\rho-1)^2}
+O(N^{-2}),
\\
\mathcal K_{\mathcal M}
-
\mathcal K_{\mathcal H}
&=
\frac12\|B\|^2
=
\frac{1}{N^2(\rho-1)^3}
+O(N^{-3}),
\end{split}
\label{eq:appC-K-above}
\end{equation}
so both normalized curvatures are $O(N^{-1})$ and positive at leading order,
while the Gauss difference is of order $N^{-2}$.

Equations~\eqref{eq:appC-h-limit},
\eqref{eq:appC-B-limit}, and
\eqref{eq:appC-K-below} summarize the geometric effect of the
boundary-to-interior saddle transition. The rank-one family retains
finite extrinsic bending on the boundary-saddle side, while for
$\rho>1$ the representative rank-two normal direction becomes
asymptotically geodesic and the intrinsic and ambient mixed tangent
curvatures agree at leading order.






\bibliographystyle{JHEP}
	\cleardoublepage

\renewcommand*{\bibname}{References}

\bibliography{references}
\end{document}